\documentclass[journal]{IEEEtran}

\IEEEoverridecommandlockouts
\makeatletter
\def\ps@headings{%
\def\@oddhead{\mbox{}\scriptsize\rightmark \hfil \thepage}%
\def\@evenhead{\scriptsize\thepage \hfil \leftmark\mbox{}}%
\def\@oddfoot{}%
\def\@evenfoot{}}
\makeatother
\usepackage{subfigure}
\usepackage{colortbl}
\usepackage{bm}

\usepackage{graphicx}  
\usepackage{url}       

\usepackage{amsmath}   
\usepackage{cite}

\usepackage{amsfonts,amssymb}
\usepackage{cite}
\usepackage{stfloats}

\usepackage{diagbox}

\usepackage{cases}
\usepackage{algorithm}
\usepackage{multirow}
\usepackage{algorithmic}
\usepackage{stfloats}
\usepackage{epstopdf}

\makeatletter

\newcommand{\Rmnum}[1]{\expandafter\@slowromancap\romannumeral #1@}
\makeatother

\UseRawInputEncoding

\newtheorem{theorem}{{Theorem}}

\newtheorem{remark}{{Remark}}

\newcommand{\ls}[1]
    {\dimen0=\fontdimen6\the\font
     \lineskip=#1\dimen0
     \advance\lineskip.5\fontdimen5\the\font
     \advance\lineskip-\dimen0
     \lineskiplimit=.9\lineskip
     \baselineskip=\lineskip
     \advance\baselineskip\dimen0
     \normallineskip\lineskip
     \normallineskiplimit\lineskiplimit
     \normalbaselineskip\baselineskip
     \ignorespaces
    }

\begin{document}

\title{Active Reconfigurable Intelligent Surface Assisted MIMO: Electromagnetic-Compliant Modeling with Mutual Coupling}
\vspace{10pt}


\author{\IEEEauthorblockN{Yang Cao, \emph{Graduate Student Member}, \emph{IEEE}, Wenchi Cheng, \emph{Senior Member}, \emph{IEEE}, Jingqing Wang, \emph{Member}, \emph{IEEE}, and Wei Zhang, \emph{Fellow}, \emph{IEEE}}\\[0.2cm]
	\vspace{-5pt}
	
	
	\vspace{-25pt}
	
	\thanks{
		
		Yang Cao, Wenchi Cheng and Jingqing Wang are with the State Key Laboratory of Integrated Services Networks, Xidian University, Xi'an,
		710071, China (e-mails: caoyang@stu.xidian.edu.cn, wccheng@xidian.edu.cn, wangjingqing00@gmail.com).
		
		Wei Zhang is with the School of Electrical Engineering and Telecommunications, University of New South Wales, Sydney,
		NSW 2052, Australia (e-mail: w.zhang@unsw.edu.au).}
}

\maketitle

\begin{abstract}
  Reconfigurable Intelligent Surfaces (RIS) represent a transformative technology for sixth-generation (6G) wireless communications, but it suffers from a significant limitation, namely the double-fading attenuation. Active RIS has emerged as a promising solution, effectively mitigating the attenuation issues associated with conventional RIS-assisted systems. However, the current academic work on active RIS focuses on the system-level optimization of active RIS, often overlooking the development of models that are compatible with its electromagnetic (EM) and physical properties. The challenge of constructing realistic, EM-compliant models for active RIS-assisted communication, as well as understanding their implications on system-level optimization, remains an open research area. To tackle these problems, in this paper we develop a novel EM-compliant model with mutual coupling (MC) for active RIS-assisted wireless systems by integrating the developed scattering-parameter ($S$-parameter) based active RIS framework with multiport network theory, which facilitates system-level analysis and optimization. To evaluate the performance of the EM-compliant active RIS model, we design the joint optimization scheme based on the transmit beamforming at the transmitter and the reflection coefficient at the active RIS to maximize the achievable rate of EM-compliant active RIS-assisted MIMO system. To tackle the inherent non-convexity of this problem, we employ the Sherman-Morrison inversion and Neumann series (SMaN)-based alternating optimization (AO) algorithm. Simulation results verified that EM property (i.e., MC effect) is an indispensable factor in the optimization process of MIMO systems. Neglecting this effect introduces a substantial performance gap, highlighting its significance in the more pronounced the MC effect is, the greater the gap in achievable rates.

\end{abstract}



\section{Introduction}
 \IEEEPARstart{A}{s} fifth-generation (5G) mobile networks become commercially widespread, potential novel technologies for sixth-generation (6G) communications are emerging in the spotlight \cite{zongshu1,tihuan1,tihuan2}. Recently, reconfigurable intelligent surface (RIS) has attracted great interest from the wireless research community and holds the promise of pioneering a new way for the evolution of next-generation communication systems \cite{tihuan3,Cao1}. RIS, a dynamic physical architecture, is composed of numerous nearly passive reflecting elements (called REs) controlled by low-cost and limited-power electronic circuits. By configuring electronic circuits, RIS permits dynamic design of wireless electromagnetic wave propagation environments to improve the performance of wireless communications. To unleash the potential of RIS, there has emerged a vast majority of research on system-level analysis and optimization of RIS \cite{dingwei,MISO1,MIMO1,NOMA}. Specifically, the authors of \cite{dingwei} investigated RIS-assisted localization of mobile user in single-input-single-output (SISO) scenario, showing that RIS can effectively enhance the accuracy of localization. The authors of \cite{MISO1} addressed the problem of maximizing the achievable average sum-rate in the RIS-assisted downlink multiuser multiple-input-single-output (MISO) system, revealing the relationship between the average sum-rate and the system parameters. The authors of \cite{MIMO1} proposed a joint optimization of the transmit precoder and the RIS elements to maximize the gain of RIS-assisted multistream multiple-input multiple-output (MIMO) communications. The authors of \cite{NOMA} proposed a solution of low complexity to balance the trade-off between achievable rate and power consumption in RIS-assisted non-orthogonal multiple access (NOMA) system.
 
 However, the RIS elements in the above examples are typically modeled as ideal scatterers, employing abstract system-level models that often fail to align with the electromagnetic (EM) properties and physical implementations. This is because they neglect important aspects of characterizing the realistic RIS, including the mutual coupling (MC) between each pair of REs at the RIS and the EM modeling of the reflection coefficients. These aspects affecting the reflection performance of the RIS must be carefully considered during the process of designing the system, which implies that oversimplification of the RIS model is impractical. Currently, some of the  published research results dedicated to bridging the gap between abstract system-level models of RIS and EM and physical implementations \cite{thus1,Renzo_2021,Shen_2022,Z_parameter,S_parameter,Z_S_parameter}. The authors of \cite{Renzo_2021} and \cite{Shen_2022} introduced EM consistency into the model of RIS by utilizing the multiport network theory, where the former relies on impedance matrices ($Z$-parameter) while the latter is based on scattering matrices ($S$-parameter). The authors of \cite{Z_parameter} further investigated MIMO systems described by RIS model based on impedance matrices and designed an optimization algorithm aimed at maximizing the sum rate of the system. In the presence of MC, the authors of \cite{S_parameter} developed the iterative algorithm to optimize the reflection coefficients of the RIS, represented based on the scattering matrix. The authors of \cite{Z_S_parameter} constructed a generalized EM consistent modeling framework for RIS demonstrating that the impedance matrix and the corresponding scattering matrix can be transformed into each other through a simple linear transformation.
 

 However, the EM consistent model for RIS with compatible EM properties and physical implementation still retains its inherent flaw, i.e., double-fading attenuation is inevitably introduced into the RIS-assisted system. Double-fading attenuation refers to the fact that the cascading channel introduced by the RIS causes the impinging signal to suffer from massive fading twice, leading to a limited performance gain of the system \cite{Active_RIS1,Active_RIScao}. As an effective solution to compensate for double-fading attenuation, the novel paradigm termed ``active RIS" has garnered significant interest among scholars in the field of wireless communications. The active RIS is uniquely constructed with each RE equipped with a reflection amplifier (RA) for phase reconstruction and amplitude amplification of weak incident signals. However, active RIS faces the same dilemma as passive RIS, i.e., the current academic efforts \cite{Active_RIS1,Active_RIS2,Active_RIS3,Active_RIS4} mainly emphasize the system-level optimization of active RIS rather than building a model that is compatible with its EM and physical characteristics. The construction of sufficiently accurate and realistic models of active RIS-assisted communications remains an open research problem. Only preliminary investigations have been conducted regarding the modeling of active RIS that aligns with compatible EM properties and physical realizations. The authors of \cite{Active_RIS_cankao} proposed a EM architecture for an active RIS based on a phase reconfigurable reflection amplifier and developed the $S$-parameter based models for its REs. The authors of \cite{Active_RIS_cankao1} designed an active RIS element, which can amplify the full polarized wave and configure the reflection phase. However, these research efforts only concentrate on the specific EM properties inherent to the active RIS elements themselves, without covering EM-consistent modeling and its implications for system-level optimization.
 
 \begin{figure*}[t]
 	\vspace{-18pt}
 	\centering
 	\includegraphics[scale=0.22]{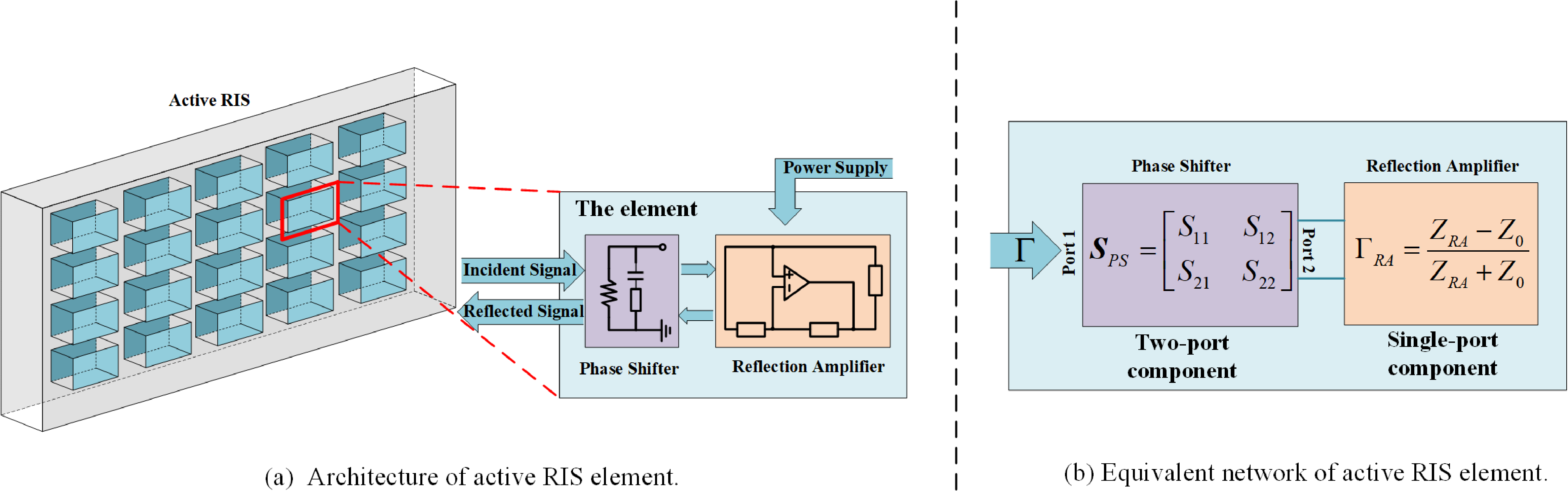}
 	\vspace{-10pt}
 	\caption{Illustration of the active RIS  architecture and the equivalent network of active RIS element.}
 	\vspace{-13pt}
 	\label{fig:shiyitu}
 \end{figure*}

 Motivated by the aforementioned challenges, the objective of this paper is to construct a novel EM-compliant model for active RIS-assisted wireless systems, which facilitates system-level analysis and optimization. Specifically, we first perform $S$-parameter based modeling of the active RIS, whose RE consists of a phase shifter (PS) and a RA in cascade. Based on the active RIS model and the multiport network theroy, we develop the EM-compliant model with MC for active RIS-assisted MIMO communication, which is effectively compatible with the EM and physical properties of active RIS. In order to evaluate the performance of the EM-compliant model with MC, we propose an optimization problem to maximize the achievable rate of the active RIS-assisted MIMO system by jointly optimizing the transmit beamforming at the BS and the reflection coefficient at the active RIS. We employ the Sherman-Morrison inversion and Neumann series (SMaN) based alternating optimization (AO) algorithm to solve the formulated non-convex problem, which is transformed into three subproblems by decoupling to seek their solutions separately. Finally, the numerical results show that EM property (i.e., MC effect) is an indispensable factor in the optimization process of MIMO systems, and optimization ignoring its effect can lead to a significant performance gap.

 The rest of this paper is organized as follows. Section~\ref{sec:System_model} introduces $S$-parameter based modeling of the active RIS and the physically consistent model for active RIS-assisted MIMO communication, which is analyzed in comparison with traditional models. In Section~\ref{sec:System_Model_Problem_formulation}, we derive the system model and formulate the non-convex optimization problem to maximize the achievable rate. In Section~\ref{sec:Alternating_optimization_perfect_CSI}, the SMaN-based AO algorithm is employed to solve the non-convex problem. In Section~\ref{sec:simulation}, the numerical results are presented to evaluate the system performance. This paper concludes with Section~\ref{sec:Conclusion}.

 \section{EM-Compliant Model and Multiport Network Analysis}
 \label{sec:System_model}

 \subsection{$S$-Parameter Based Modeling of Active RIS}
 \label{subsec:EM_Active_RIS}
 
 Active RIS has recently gained attention as a technology, which effectively compensates for the double-fading attenuation introduced by passive RIS in communication systems. Each RE of the active RIS is equipped with an additional amplifier to perform the function of adjusting the phase and amplitude of the incident signal. As shown in Fig.~\ref{fig:shiyitu}{(a)}, the RE of the active RIS consists of a PS and a RA, where the PS reconfigures the phase and the RA amplifies the incident signal by utilizing transistors. Based on multiport network theory \cite{Ivrlac_Nossek_2010}, the equivalent network of the RE at the active RIS is modeled as shown in Fig.~\ref{fig:shiyitu}{(b)}. Specifically, the PS is modeled as a two-port component, which is characterized by employing an $S$-parameter matrix, denoted by $\boldsymbol{\rm S}_{P\hspace{-0.3mm}S}$, as follows:
 \begin{align}\label{S_ps}
 \boldsymbol{\rm S}_{P\hspace{-0.3mm}S}=\begin{bmatrix}
 S_{11} & S_{12} \\
 S_{21} & S_{22}
 \end{bmatrix},
 \end{align}
 where $S_{11}$ and $S_{22}$ represent the reflection coefficients of the corresponding ports, and $S_{12}$ and $S_{21}$ are the forward and reverse losses. Then, since the series connection of the PS and the RA constitutes the RE of the active RIS, the RA can be modeled as a load connected to the port $2$ of the PS. The RA has only one port as input and output, which enables the function of amplifying the reflecting signal. The reflection coefficient of the RA can be described as
 \begin{equation}\label{gamma_RA}
 {\Gamma}_{R\hspace{-0.3mm}A} = \frac{Z_{R\hspace{-0.3mm}A}-Z_0}{Z_{R\hspace{-0.3mm}A}+Z_0}=\alpha_{R\hspace{-0.3mm}A} e^{j\theta_{R\hspace{-0.3mm}A}},
 \end{equation}
 where $Z_0=50 \Omega$, $Z_{R\hspace{-0.3mm}A}$, $\alpha_{R\hspace{-0.3mm}A}$, and $\theta_{R\hspace{-0.3mm}A}$ denote the characteristic impedance, the input impedance, the amplitude, and the phase of the RA, respectively. Therefore, according to Mason rules \cite{active_model1}, the reflection coefficient of the RE is derived as follows:
 \begin{align}
 {\Gamma} = S_{11}+\frac{S_{12}S_{21}{\Gamma}_{R\hspace{-0.3mm}A}}{1-S_{22}{\Gamma}_{R\hspace{-0.3mm}A}}.
 \end{align}
 
  \begin{figure}[t]
 	\vspace{-10pt}
 	\centering
 	\includegraphics[scale=0.075]{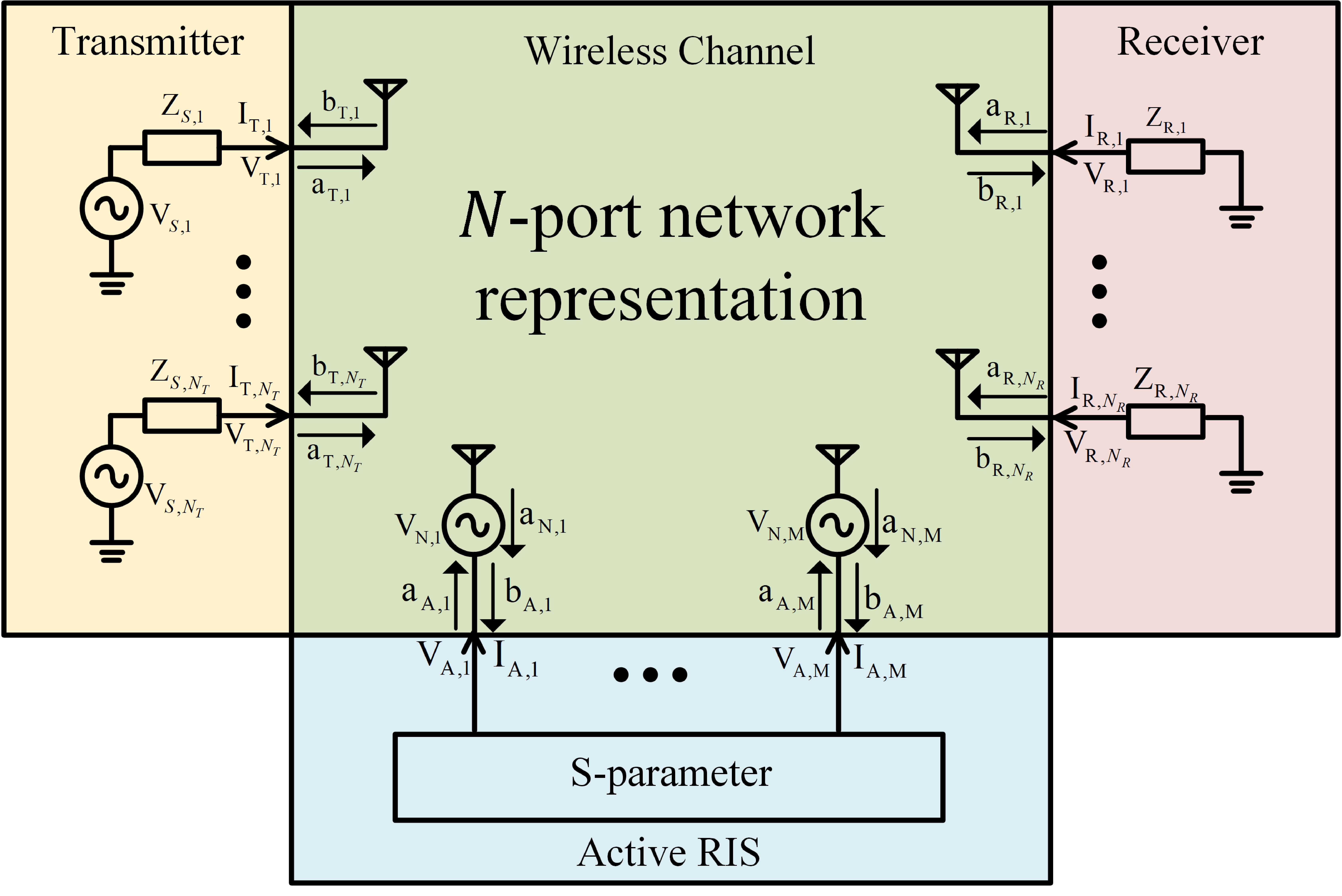}
 	\caption{Illustration of the EM-compliant model based on multiport network theory.}
 	\vspace{-13pt}
 	\label{fig:multiple_port}
 \end{figure}
 
 In fact, the complexity of the $\Gamma$ construction necessitates further approximate simplification. Since the reflection coefficient $S_{11}$ of the PS is typically low (less than $-15$ dB \cite{active_model2}) and considering that $\Gamma$ must be greater than 1, then the reflection coefficient of the RE can be further simplified by neglecting $S_{11}$ as follows:
 \begin{align}\label{gamma_1}
 {\Gamma} \approx\frac{S_{12}S_{21}{\Gamma}_{R\hspace{-0.3mm}A}}{1-S_{22}{\Gamma}_{R\hspace{-0.3mm}A}}.
 \end{align}
 Furthermore, balancing gain and stability are critical in designing RE for active RIS \cite{active_model1}. In order to achieve a stable and significant reflection gain, the numerator and denominator of Eq.~(\ref{gamma_1}) are required to maintain stability at different phase conditions. To maintain a stable denominator, it is essential that $S_{22}{\Gamma}_{R\hspace{-0.3mm}A}$ approaches zero in the design criteria \cite{Active_RIS_cankao}. And for the numerator, we further define
 $S_{12}=S_{21}=L_{P\hspace{-0.3mm}S}e^{j\theta}$, where $\theta$ and $L_{PS}$ represent the phase and the insertion loss of the PS \cite{active_model3}. The insertion loss is considered to remain constant to ensure numerator stabilization. Also, we further ignore the slight change in the phase of RA, ie., $\theta_{R\hspace{-0.3mm}A}=0$. Finally, the reflection coefficient of the RE is simplified as follows:
  \begin{align}\label{gamma_2}
 {\Gamma}\approx L_{P\hspace{-0.3mm}S}^{2}\alpha_{R\hspace{-0.3mm}A}e^{j2\theta}.
 \end{align}
 Based on the above design criteria, the active RIS is capable of achieving stable and significant system gain by adjusting the phase of the PS and amplitude of the RA. It is evident that the model of the active RIS differs from that of the conventional passive RIS \cite{Cao1}, specifically the overall gain due to the active RIS is the gain of the RA minus twice the insertion loss and the corresponding phase is twice the phase of the PS.

 \subsection{EM-Compliant Model Based on Multiport Network Theory}
 \label{sec:system_Model}

 In this paper, we consider an active RIS-assisted MIMO system, where a transmitter equipped with $N_T$ antennas, communicates to a receiver equipped with $N_R$ antennas with the aid of an active RIS with $M$ REs. Let the set of antennas at the transmitter, the set of antennas at the receiver, and the set of REs at the active RIS be defined by $\mathcal{N}_T=\{1,2,\cdots,N_T\}$, $\mathcal{N}_R=\{1,2,\cdots,N_R\}$, and $\mathcal{M}=\{1,2,\cdots,M\}$. For further analysis, the antennas on the transmitter and receiver as well as REs of the active RIS are modeled as ports of a $N$-port network, respectively, where $N = N_T+M+N_R$. As shown in Fig.~\ref{fig:multiple_port}, we take into account the employment of $S$-parameter for modeling the relationship between the transmitter and the receiver. Based on the multiport network theory, the $N$-port network can be modeled by its scattering matrix $\boldsymbol{\rm S}\in\mathbb{C}^{N\times N}$, as follows:
 \begin{align}\label{S_parameter}
 \boldsymbol{\rm b}= \boldsymbol{\rm S}\boldsymbol{\rm a},
 \end{align}
 where $\boldsymbol{\rm b}\in\mathbb{C}^{N\times 1}$ and $\boldsymbol{\rm a}\in\mathbb{C}^{N\times 1}$ represent scattered and incident voltage waves. Based on the block matrix division, we can obtain
 \begin{align} \label{S_parameter1}
 &\boldsymbol{\rm a}= \left[\boldsymbol{\rm a}_T^T, \boldsymbol{\rm a}_A^T, \boldsymbol{\rm a}_R^T\right]^T,\\
 &\boldsymbol{\rm S}=\begin{bmatrix}
 \boldsymbol{\rm S}_{TT} & \boldsymbol{\rm S}_{TA} & \boldsymbol{\rm S}_{TR}\\
 \boldsymbol{\rm S}_{AT} & \boldsymbol{\rm S}_{AA} & \boldsymbol{\rm S}_{AR}\\
 \boldsymbol{\rm S}_{RT} & \boldsymbol{\rm S}_{RA} & \boldsymbol{\rm S}_{RR}\label{S_parameter2}\\
 \end{bmatrix},\\
 &\boldsymbol{\rm b}= \left[\boldsymbol{\rm b}_T^T, \boldsymbol{\rm b}_A^T, \boldsymbol{\rm b}_R^T\right]^T,\label{S_parameter3}
 \end{align}
 where $\boldsymbol{\rm a}_\chi$ and $\boldsymbol{\rm b}_\chi$ for $\chi\in\left\{T,A,R\right\}$ denote the incident and scattered voltage waves at the antennas (or REs) of the transmitter, active RIS, and receiver, respectively. The matrix $\boldsymbol{\rm S}_{\chi_1\chi_2}$, where $\chi_1, \chi_2\in\left\{T,A,R\right\}$, is the scattering sub-matrix between the port of $\chi_1$ and $\chi_2$. In addition, the voltage waves are associated with the voltage $\boldsymbol{\rm v}$ and current $\boldsymbol{\rm i}$ of the $N$-port network, respectively, as follows \cite{Davis_Stutzman_2005}:
 \begin{align}
  \boldsymbol{\rm v}=\sqrt{Z_{0}}(\boldsymbol{\rm a}+\boldsymbol{\rm b}),~~ \boldsymbol{\rm i}=\frac{\boldsymbol{\rm a}-\boldsymbol{\rm b}}{\sqrt{Z_{0}}}.
 \end{align}
 
 For the transmitter, each of its $N_T$ ports is connected to a voltage generator, so the relationship between $\boldsymbol{\rm a}_T$ and $\boldsymbol{\rm b}_T$ can be constructed as
 \begin{align}\label{a_T}
 \boldsymbol{\rm a}_T = \boldsymbol{\rm a}_S+\boldsymbol{\rm \Gamma}_{T}\boldsymbol{\rm b}_T,
 \end{align}
 where $\boldsymbol{\rm a}_S$ and $\boldsymbol{\rm \Gamma}_{T}\in\mathbb{C}^{N_T\times N_T}$ are the voltage source wave corresponding to the voltage generator and the diagonal matrix of reflection coefficients at the transmitter, where its $(n_T, n_T)$th element is the reflection coefficient of the $n_T$th voltage source impedance $Z_{T,n_T}$. Thus, the relationship between $\boldsymbol{\rm \Gamma}_{T}$ and the voltage source impedance $\boldsymbol{\rm Z}_{T}$ is given by $\boldsymbol{\rm \Gamma}_{T}=\left(\boldsymbol{\rm Z}_{T}+Z_0\boldsymbol{\rm I}_{N_T}\right)^{-1}\left(\boldsymbol{\rm Z}_{T}-Z_0\boldsymbol{\rm I}_{N_T}\right)$. Besides, the relationship between $\boldsymbol{\rm a}_R$ and $\boldsymbol{\rm b}_R$ at the receiver can be modeled as
 \begin{align}\label{a_R}
 \boldsymbol{\rm a}_R = \boldsymbol{\rm \Gamma}_{R}\boldsymbol{\rm b}_R,
 \end{align}
 where $\boldsymbol{\rm \Gamma}_{T}\in\mathbb{C}^{N_T\times N_T}$ is the diagonal matrix of reflection coefficients at the receiver and there exists the relationship $\boldsymbol{\rm \Gamma}_{R}=\left(\boldsymbol{\rm Z}_{R}+Z_0\boldsymbol{\rm I}_{N_R}\right)^{-1}\left(\boldsymbol{\rm Z}_{R}-Z_0\boldsymbol{\rm I}_{N_R}\right)$ with $\boldsymbol{\rm Z}_{R}$ being the load impedance connected with the receiver.
 
 In contrast to traditional passive RIS, active RIS amplification can not only act on the desired signal, but also the thermal noise at the active RIS. Therefore, we abstract and relocate the thermal noise sources within the active RIS to its ports, while treating the intrinsic part of the active RIS as a noise-free ideal device. Let the thermal noise source wave be defined as $\boldsymbol{\rm a}_N$. Based on above analysis, the relationship between $\boldsymbol{\rm a}_A$ and $\boldsymbol{\rm b}_A$ at the active RIS can be expressed as follows:
 \begin{align}\label{a_A}
 \boldsymbol{\rm a}_A = \boldsymbol{\rm \Gamma}_{A}\boldsymbol{\rm a}_N +\boldsymbol{\rm \Gamma}_{A}\boldsymbol{\rm b}_A,
 \end{align}
 where $\boldsymbol{\rm \Gamma}_A$ denotes the reflection coefficients at the active RIS. In addition, $\boldsymbol{\rm \Gamma}_A = \mbox{Diag}\left(\left[\Gamma_{A,1},\cdots,\Gamma_{A,M}\right]^T\right)$ is a diagonal matrix, where the specific expression of $\Gamma_{A,m}, \forall m\in\mathcal{M}$ is given in Eq.~(\ref{gamma_2}) and $\mbox{Diag}\left(\cdot\right)$ refers to a diagonal matrix. Thus, $\boldsymbol{\rm \Gamma}_A$ can be rewritten as 
 \begin{align}\label{gamma_3}
 \boldsymbol{\rm \Gamma}_A=L_{P\hspace{-0.3mm}S}^{2}\boldsymbol{\rm \Lambda}_{R\hspace{-0.3mm}A}\exp\left(j2\boldsymbol{\rm \Theta}\right),
 \end{align}
 where $\boldsymbol{\rm \Lambda}_{R\hspace{-0.3mm}A}=\mbox{Diag}\left(\left[\alpha_{R\hspace{-0.3mm}A,1},\cdots,\alpha_{R\hspace{-0.3mm}A,M}\right]^T\right)$ and $\boldsymbol{\rm \Theta}=\mbox{Diag}\left(\left[\theta_1,\cdots,\theta_M\right]^T\right)$ are the diagonal amplitude matrix and the diagonal phase matrix, respectively. Besides, $\alpha_{R\hspace{-0.3mm}A,m}$ and $\theta_m$ represent the amplitude of RA and the phase of PS at the $m$th RE of active RIS, respectively.

 Based on the above heuristic results for the $S$-parameter, we introduce an EM-compliant model for active RIS-assisted communication, which takes into account any number of coupled antenna elements at the transmitter and receiver, as well as the REs at the RIS. The EM-compliant model is given in the following theorem.
 \begin{theorem}\label{theorem_channnel}
 	Based on multiport network theory, we formulate $\boldsymbol{\rm b}_R$ as a function of $\boldsymbol{\rm a}_S$ and $\boldsymbol{\rm a}_N$ to derive the EM-compliant model for active RIS-assisted MIMO communication as follows: 
 	\begin{align}\label{channel_model1}
 	\begin{split}
 	\boldsymbol{\rm b}_R =& \left(\boldsymbol{\rm I}_{N_R}-\boldsymbol{\rm \widetilde S}_{RR}\boldsymbol{\rm \Gamma}_{R}\right)^{-1}\boldsymbol{\rm \widetilde S}_{RT}\left(\boldsymbol{\rm I}_{N_T}-\boldsymbol{\rm \Gamma}_{T}\boldsymbol{\rm \widehat S}_{TT}\right)^{-1}\\
 	&\qquad\qquad\qquad\times\left(\boldsymbol{\rm a}_S+\boldsymbol{\rm \Gamma}_{T}\boldsymbol{\rm \widehat S}_{N}\boldsymbol{\rm a}_N\right)+\boldsymbol{\rm S}_{N}\boldsymbol{\rm a}_N,
 	\end{split}
 	\end{align}
 	where we introduce the definitions as follows:
 	\begin{align}
 	&\boldsymbol{\rm \widetilde S}_{\chi_1\chi_2}\triangleq\boldsymbol{\rm S}_{\chi_1\chi_2}+\boldsymbol{\rm  S}_{\chi_1A}\left(\boldsymbol{\rm I}_{M}-\boldsymbol{\rm \Gamma}_{A}\boldsymbol{\rm  S}_{AA}\right)^{-1}\boldsymbol{\rm \Gamma}_{A}\boldsymbol{\rm S}_{A\chi_2},\notag\\
 	&\boldsymbol{\rm \widehat S}_{TT}\triangleq\boldsymbol{\rm \widetilde S}_{TT}+\boldsymbol{\rm \widetilde S}_{TR}\boldsymbol{\rm \widetilde S}_{RR}^{-1}\left(\boldsymbol{\rm I}_{N_R}-\boldsymbol{\rm \widetilde S}_{RR}\boldsymbol{\rm \Gamma}_{R}\right)^{-1}\boldsymbol{\rm \widetilde S}_{RT},\notag\\
 	&\boldsymbol{\rm S}_{N}\triangleq\left(\boldsymbol{\rm I}_{N_R}-\boldsymbol{\rm \widetilde S}_{RR}\boldsymbol{\rm \Gamma}_{R}\right)^{-1}\boldsymbol{\rm S}_{RA}\boldsymbol{\rm \Gamma}_{A}\left(\boldsymbol{\rm I}_{M}-\boldsymbol{\rm \Gamma}_{A}\boldsymbol{\rm S}_{AA}\right)^{-1}\hspace{-2mm},\notag\\
 	&\boldsymbol{\rm \widetilde S}_{N}\triangleq\boldsymbol{\rm S}_{N}-\boldsymbol{\rm S}_{RA}\boldsymbol{\rm \Gamma}_{A}\left(\boldsymbol{\rm I}_{M}-\boldsymbol{\rm \Gamma}_{A}\boldsymbol{\rm S}_{AA}\right)^{-1},\notag\\
 	&\boldsymbol{\rm \widehat S}_{N}\triangleq\boldsymbol{\rm \widetilde S}_{TR}\boldsymbol{\rm \widetilde S}_{RR}^{-1}\boldsymbol{\rm \widetilde S}_{N}+\boldsymbol{\rm S}_{TA}\boldsymbol{\rm \Gamma}_{A}\left(\boldsymbol{\rm I}_{M}-\boldsymbol{\rm \Gamma}_{A}\boldsymbol{\rm S}_{AA}\right)^{-1},
 	\end{align}
 	with $\chi_1, \chi_2\in\left\{T,R\right\}$ and $\boldsymbol{\rm I}$ being the identity matrix of the dimension corresponding to the subscripts.	
 \end{theorem}
 \textit{proof:} Please see Appendix A.$\hfill\blacksquare$
 
 \begin{remark}
 	Inspired by the EM-consistent passive RIS model derived in \cite{S_parameter}, we develop the first EM-compliant model for active RIS-assisted MIMO systems, which provides a clear and detailed understanding of the effects of impedance mismatch and MC of the transmitter, active RIS, and receiver. A key contribution of this model is its rigorous inclusion of the impact of thermal noise, which is amplified by the RA during active RIS operation. However, while the EM-compliant model for active RIS-assisted MIMO systems offers a comprehensive description, it results in highly complex expressions. These expressions, while essential for an in-depth analysis, must be simplified for practical system-level analysis and optimization.
 \end{remark}
  
 To gain insight into the role of the active RIS in the communication model, we simplify the expression by considering a special case,  where the source impedance at the transmitter and the load impedance at the receiver are both reference impedances $Z_0$ such that $\boldsymbol{\rm \Gamma}_{T} = 0$ and $\boldsymbol{\rm \Gamma}_{R} = 0$. Thus, the EM-compliant model with MC for active RIS-assisted MIMO communication can be reduced as follows:
 \begin{align}\label{channel_model2}
 \begin{split}
 \boldsymbol{\rm b}_R =\boldsymbol{\rm H}_{S}^{e}\boldsymbol{\rm a}_S+\boldsymbol{\rm H}_{S}^{n}\boldsymbol{\rm a}_N,
 \end{split}
 \end{align}
 where we define
 \begin{align}\label{Hs}
 \boldsymbol{\rm H}_{S}^{e}&\triangleq\boldsymbol{\rm S}_{RT}+\boldsymbol{\rm  S}_{RA}\left(\boldsymbol{\rm I}_{M}-\boldsymbol{\rm \Gamma}_{A}\boldsymbol{\rm  S}_{AA}\right)^{-1}\boldsymbol{\rm \Gamma}_{A}\boldsymbol{\rm S}_{AT},\\
 \boldsymbol{\rm H}_{S}^{n}&\triangleq\boldsymbol{\rm  S}_{RA}\left(\boldsymbol{\rm I}_{M}-\boldsymbol{\rm \Gamma}_{A}\boldsymbol{\rm  S}_{AA}\right)^{-1}\boldsymbol{\rm \Gamma}_{A}.\label{He}
 \end{align}
 
 \subsection{Differences From Conventional Model}
 \label{conventional_model}
 
 According to \cite{Dinkelbach}, the conventional model (scattering matrix) of active RIS-assisted MIMO utilized in communication theory is generally indicated as follows:
 \begin{align}\label{channel_model3}
 \begin{split}
 \boldsymbol{\rm b}_R =\left(\boldsymbol{\rm S}_{RT}+\boldsymbol{\rm  S}_{RA}\boldsymbol{\rm \Gamma}_{A}\boldsymbol{\rm S}_{AT}\right)\boldsymbol{\rm a}_S+\boldsymbol{\rm  S}_{RA}\boldsymbol{\rm \Gamma}_{A}\boldsymbol{\rm a}_N.
 \end{split}
 \end{align}
 Then, we let $\boldsymbol{\rm H}_{C}^{e}=\boldsymbol{\rm S}_{RT}+\boldsymbol{\rm  S}_{RA}\boldsymbol{\rm \Gamma}_{A}\boldsymbol{\rm S}_{AT}$ and $\boldsymbol{\rm H}_{C}^{n}=\boldsymbol{\rm  S}_{RA}\boldsymbol{\rm \Gamma}_{A}$. Comparing $\boldsymbol{\rm H}_{S}^{e}$ and $\boldsymbol{\rm H}_{S}^{n}$ with $\boldsymbol{\rm H}_{C}^{e}$ and $\boldsymbol{\rm H}_{C}^{n}$ separately, some basic differences can be distinguished. Traditional active RIS model implicitly ignores the MC between REs at the active RIS. We can see that if the matrices $\boldsymbol{\rm S}_{AA}$ indicating MC are equal to $0$, $\boldsymbol{\rm H}_{S}^{e}$ = $\boldsymbol{\rm H}_{C}^{e}$ and $\boldsymbol{\rm H}_{S}^{n}$ = $\boldsymbol{\rm H}_{C}^{n}$. 
 
 In addition, the conventional active RIS model satisfies the condition $\boldsymbol{\rm S}_{AA}=0$, which also states that the self-impedance of the REs is equal to the reference impedance. Furthermore, based on \cite{BD_RIS}, it is well known that $\boldsymbol{\rm S}_{RT}$ in $\boldsymbol{\rm H}_{S}^{e}$ and $\boldsymbol{\rm H}_{C}^{e}$ are different. That is, when the direct link between the transmitter and receiver is blocked, there is yet a structural scattering part of $\boldsymbol{\rm S}_{RT}$ in $\boldsymbol{\rm H}_{S}^{e}$, while $\boldsymbol{\rm S}_{RT}=0$ in $\boldsymbol{\rm H}_{C}^{e}$. This indicates that the structural scattering of the active RIS is completely ignored in $\boldsymbol{\rm H}_{C}^{e}$, while it is carefully considered in $\boldsymbol{\rm H}_{S}^{e}$. Thus, our derived EM-compliant active RIS model with MC incorporates the conventional active RIS model without MC. By ignoring MC, the former and the latter can be made equivalent. More importantly, the primary feature that distinguishes the two above models is that the system channel response is nonlinear when the EM-compliant model with MC for active RIS-assisted MIMO system is utilized as a function of the reflection coefficient at the active RIS. This profoundly reveals the effect of electromagnetic MC among the dense radiating elements of the active RIS in practice.

 Besides, the EM-compliant model with MC for passive RIS-assisted communication in \cite{Physically_Consistent} can be formulated as
 \begin{align}\label{channel_model4}
 \begin{split}
 \boldsymbol{\rm b}_R &=\left(\boldsymbol{\rm S}_{RT}+\boldsymbol{\rm  S}_{RP}\left(\boldsymbol{\rm I}_{M}-\boldsymbol{\rm \Gamma}_{P}\boldsymbol{\rm  S}_{PP}\right)^{-1}\boldsymbol{\rm \Gamma}_{P}\boldsymbol{\rm S}_{PT}\right)\boldsymbol{\rm a}_S\\
 &=\boldsymbol{\rm H}_{P}^{e}\boldsymbol{\rm a}_S,
 \end{split}
 \end{align}
  where the subscript $P$ stands for passive RIS with its associated parameters defined similarly to those in the active RIS model. Comparing Eqs.~(\ref{channel_model2}) and (\ref{channel_model4}), it can be seen that the thermal noise term at the RIS is neglected in the EM-compliant model of the passive RIS without considering the specific setup conditions of the reflection coefficients of the active and passive RIS. This is reasonable because thermal noise in passive RIS is small enough to be negligible, whereas on the contrary, thermal noise in active RIS is amplified by the RA and thereby difficult to ignore. 
  
   \section{System Model And Problem Formulation}
  \label{sec:System_Model_Problem_formulation}
  
 \subsection{System Model}
 \label{sec:System_Model}
 
  Based on the EM-compliant model with MC for active RIS, we develop the active RIS-assisted MIMO system model. The transmitted signal vector, denoted by $\boldsymbol{\rm {x}}$, at the transmitter is expressed as follows:
  \begin{align}
  \boldsymbol{\rm {x}} = \boldsymbol{\rm {W}}\boldsymbol{\rm {s}},
  \end{align}
  where $\boldsymbol{\rm {W}}\in \mathbb{C}^{N_T\times N_R}$ and $\boldsymbol{\rm {s}}\in \mathbb{C}^{N_R\times 1}$ denote the transmit beamforming matrix and the desired signal, respectively,
  where $\boldsymbol{\rm {s}}\sim\mathcal{CN}(0,\boldsymbol{\rm {I}}_{N_R})$ is modeled as a zero mean and unit variance random matrix with independent and identically distributed (i.i.d.). In addition, $\boldsymbol{\rm {W}}$ satisfies the maximum transmit power constraint: $\mbox{Tr}\left(\boldsymbol{\rm W}\boldsymbol{\rm W}^H\right)\le P_{\max}$, where $\mbox{Tr}\left(\cdot\right)$ and $P_{\max}$ represent the trace of the matrix and the maximum transmit power budget.
  
  Based on the EM-compliant model for active RIS with MC, the signal received by the receiver can be constructed as follows:
  \begin{align}
  \boldsymbol{\rm {y}} = \boldsymbol{\rm {H}}_{S}^{e}\boldsymbol{\rm {x}}+\boldsymbol{\rm {H}}_{S}^{n}\boldsymbol{\rm n}_{\scriptscriptstyle R\hspace{-0.2mm}I\hspace{-0.2mm}S}+\boldsymbol{\rm n},
  \end{align}
  where $\boldsymbol{\rm n}_{\scriptscriptstyle R\hspace{-0.2mm}I\hspace{-0.2mm}S}\sim\mathcal{CN}(0,\sigma^2_{\scriptscriptstyle R\hspace{-0.2mm}I\hspace{-0.2mm}S}\boldsymbol{\rm I}_M)$ and $\boldsymbol{\rm n}\sim\mathcal{CN}(0,\sigma^2\boldsymbol{\rm I}_M)$ denote the additive complex Gaussian noises at the active RIS and the receiver, respectively. Then, we can derive the achievable rate of the active RIS-assisted MIMO system, as follows:
  \begin{align}
  \begin{split}
  R\left(\boldsymbol{\rm W},\boldsymbol{\rm \Gamma}_A\right)\hspace{-1mm}=\hspace{-1mm}\log_2\det\hspace{-1mm}\left(\hspace{-1mm}\boldsymbol{\rm I}_{N_R}\hspace{-1mm}+\hspace{-1mm}\frac{\boldsymbol{\rm H}_{S}^e\boldsymbol{\rm W}\boldsymbol{\rm W}^H(\boldsymbol{\rm H}_{S}^e)^H}{\sigma^2_{\scriptscriptstyle R\hspace{-0.2mm}I\hspace{-0.2mm}S}\boldsymbol{\rm H}_{S}^n(\boldsymbol{\rm H}_{S}^n)^H\hspace{-1mm}+\hspace{-1mm}\sigma^2\boldsymbol{\rm I}_{N_R}}\hspace{-1mm}\right)\hspace{-1mm}
  ,
  \end{split}
  \end{align}
  where $\det\left(\cdot\right)$ and $(\cdot)^H$ represent the determinant of a matrix and the conjugate transpose.
  
  In addition, since the power budget for signal amplification at the active RIS is limited, it is imperative to conduct an analysis of the amplification power in the EM-compliant active RIS model. Based on the multiport network theory, we can derive the output wave at the EM-compliant active RIS in the following theorem.
  
  \begin{theorem}\label{theorem_2}
  	Taking into account $\boldsymbol{\rm \Gamma}_{T} = 0$, $\boldsymbol{\rm \Gamma}_{R} = 0$, and the unilateral approximation ($\boldsymbol{\rm S}_{TA} = 0$, $\boldsymbol{\rm S}_{TR} = 0$, and $\boldsymbol{\rm S}_{AR} = 0$) \cite{Ivrlac_Nossek_2010}, we derive that the output wave $\boldsymbol{\rm a}_A$ at the active RIS can be formulated as a function of $\boldsymbol{\rm a}_S$ and $\boldsymbol{\rm a}_N$ as follows: 
  	\begin{align}\label{theorem_21}
  	\begin{split}
  	\boldsymbol{\rm a}_A =\boldsymbol{\rm \breve {H}}_{S}^{e}\boldsymbol{\rm a}_S+\boldsymbol{\rm \breve {H}}_{S}^{n}\boldsymbol{\rm a}_N,
  	\end{split}
  	\end{align}
  	where we introduce the definition as follows:
  	\begin{align}\label{H_A}
  	&\boldsymbol{\rm \breve {H}}_{S}^{e}\triangleq\left(\boldsymbol{\rm I}_{M}-\boldsymbol{\rm \Gamma}_{A}\boldsymbol{\rm  S}_{AA}\right)^{-1}\boldsymbol{\rm \Gamma}_{A}\boldsymbol{\rm S}_{AT},\\
  	&\boldsymbol{\rm \breve {H}}_{S}^{n}\triangleq\left(\boldsymbol{\rm I}_{M}-\boldsymbol{\rm \Gamma}_{A}\boldsymbol{\rm  S}_{AA}\right)^{-1}\boldsymbol{\rm \Gamma}_{A}.\label{H_B}
  	\end{align}
  \end{theorem}
  \textit{proof:} Please see Appendix B.$\hfill\blacksquare$ 
  
  Therefore, based on Theorem~\ref{theorem_2}, the amplification power constraint at the active RIS is considered as follows:
  \begin{align}\label{power_constraint}
  \begin{split}
  \mbox{Tr}\hspace{-1mm}\left(\hspace{-1mm}\boldsymbol{\rm \breve {H}}_{S}^{e}\boldsymbol{\rm W}\boldsymbol{\rm W}^H\hspace{-1mm}\left(\boldsymbol{\rm \breve {H}}_{S}^{e}\right)^H\hspace{-0.8mm}\right)\hspace{-1mm}+\hspace{-1mm}\sigma^2_{\scriptscriptstyle R\hspace{-0.2mm}I\hspace{-0.2mm}S}\mbox{Tr}\hspace{-1mm}\left(\hspace{-1mm}\boldsymbol{\rm \breve {H}}_{S}^{n}\left(\boldsymbol{\rm \breve {H}}_{S}^{n}\right)^H\hspace{-0.8mm}\right)\hspace{-1mm}\le\hspace{-1mm} P_{\max}^A,
  \end{split}
  \end{align}
  where $P_{\max}^A$ denotes the amplification
  power budget at the active RIS.
  
 \subsection{Problem Formulation And Transformation}
 \label{sec:Problem_formulation_1}
 In the subsection, we formulate an optimization problem to maximize the system achievable rate by jointly optimizing the transmit beamforming of the transmitter and the reflection coefficient matrix of the active RIS, subject to maximum transmit power and maximum amplitude of the reflection coefficient. As such, the optimization problem is described as follows:
 \begin{subequations}\label{Problem:P1}
	\begin{alignat}{2}
	\textbf{\textit{P}1:}
	&\mathop{\max}\limits_{\boldsymbol{\rm W},\boldsymbol{\rm \Gamma}_A} R\left(\boldsymbol{\rm W},\boldsymbol{\rm \Gamma}_A\right)  \notag \\
	{\rm{s.t.}}:&1).\ \mbox{Tr}\left(\boldsymbol{\rm W}\boldsymbol{\rm W}^H\right)\le P_{\max};\\
	&2).\  1\le\left\vert\Gamma_{A,m}\right\vert \le \Gamma_{\max}, \forall m \in \mathcal{M};\\
	&3).\ (\ref{power_constraint})\notag,
	\end{alignat}
\end{subequations}
 where $\Gamma_{\max}$ denotes the maximum amplitude of the reflection coefficient. Note that the optimization problem $\textbf{\textit{P}1}$ is a non-convex problem due to the fact that the optimization variables ($\boldsymbol{\rm W}$ and $\boldsymbol{\rm \Gamma}_A$) are coupled in the objective function of $\textbf{\textit{P}1}$. In addition, the optimization problem is generally intractable owing to the existence of fractional forms and a large number of matrix inverses in the objective function.
 
 To simplify the optimization problem, the achievable rate problem with determinant can be transformed into an equivalent mean square error (MSE) estimation problem by introducing auxiliary variables \cite{MSE_1}. For the signal received at the receiver, we introduce the detection matrix $\boldsymbol{\rm D}\in\mathbb{C}^{N_R\times N_R}$. The estimated signal vector, denoted by $\boldsymbol{\rm \widehat{s}}$, at the receiver can be expressed as follows:
 \begin{align}
 \boldsymbol{\rm \widehat{s}}=\boldsymbol{\rm D}^H\boldsymbol{\rm y}=\boldsymbol{\rm D}^H\left(\boldsymbol{\rm {H}}_{S}^{e}\boldsymbol{\rm {x}}+\boldsymbol{\rm {H}}_{S}^{n}\boldsymbol{\rm n}_{\scriptscriptstyle R\hspace{-0.2mm}I\hspace{-0.2mm}S}+\boldsymbol{\rm n}\right).
 \end{align}
 Thus, the MSE matrix between the estimated and received signals at the receiver, denoted by $\boldsymbol{\rm U}$, is derived as
 \begin{align}\label{Problem:P0.5}
 \begin{split}
 \boldsymbol{\rm U}=&\mathbb{E}\left[\left(\boldsymbol{\rm \widehat{s}}-\boldsymbol{\rm {s}}\right)\left(\boldsymbol{\rm \widehat{s}}-\boldsymbol{\rm {s}}\right)^H\right]\\
 =&\left(\boldsymbol{\rm D}^H\boldsymbol{\rm {H}}_{S}^{e}\boldsymbol{\rm {W}}-\boldsymbol{\rm {I}}_{N_R}\right)\hspace{-1mm}\left(\boldsymbol{\rm D}^H\boldsymbol{\rm {H}}_{S}^{e}\boldsymbol{\rm {W}}-\boldsymbol{\rm {I}}_{N_R}\right)^H\hspace{-3mm}+\boldsymbol{\rm D}^H\boldsymbol{\rm N}\boldsymbol{\rm D},
 \end{split}
 \end{align}
 where $\boldsymbol{\rm N}=\sigma^2_{\scriptscriptstyle R\hspace{-0.2mm}I\hspace{-0.2mm}S}\boldsymbol{\rm H}_{S}^n(\boldsymbol{\rm H}_{S}^n)^H\hspace{-1mm}+\hspace{-1mm}\sigma^2\boldsymbol{\rm I}_{N_R}$ represents the covariance matrix of the noise in the received signal.
 
 In the following, we reformulate the objective function in $\textbf{\textit{P}1}$ by introducing auxiliary variable $\boldsymbol{\rm V}\in\mathbb{C}^{N_R\times N_R}$, $\boldsymbol{\rm V}\succ 0$, as follows:
 \begin{align}\label{Problem:P1.5}
 \begin{split}
 R\left(\boldsymbol{\rm W},\boldsymbol{\rm \Gamma}_A\right)=\max\limits_{\boldsymbol{\rm D},\boldsymbol{\rm V}\succ 0}\log_2\det\left(\boldsymbol{\rm V}\right)-\mbox{Tr}\left(\boldsymbol{\rm V}\boldsymbol{\rm U}\right)+\mbox{Tr}\left(\boldsymbol{\rm I}_{N_R}\right) .
 \end{split}
 \end{align}
 Based on Eqs.~(\ref{Problem:P0.5}) and (\ref{Problem:P1.5}), it can be noted that $R$ is a concave function with respect to the variables $\boldsymbol{\rm D}$ and $\boldsymbol{\rm V}$ \cite{MSE_1}. Thus, let $\frac{\partial R}{\partial \boldsymbol{\rm D}}=0$, we can obtain the optimal value of the variable $\boldsymbol{\rm D}$, as follows:
 \begin{align}\label{para_D}
 \boldsymbol{\rm D}^{\star}=\left(\boldsymbol{\rm H}_{S}^e\boldsymbol{\rm W}\boldsymbol{\rm W}^H(\boldsymbol{\rm H}_{S}^e)^H+\boldsymbol{\rm N}\right)^{-1}\boldsymbol{\rm H}_{S}^e\boldsymbol{\rm W}.
 \end{align}
 Substituting $\boldsymbol{\rm D}^{\star}$ into Eq.~(\ref{Problem:P1.5}) and then letting $\frac{\partial R}{\partial \boldsymbol{\rm V}}=0$, the optimal value of $\boldsymbol{\rm V}$ can be obtained as follows:
 \begin{align}\label{para_V}
 \boldsymbol{\rm V}^{\star}=\left(\boldsymbol{\rm U}\left(\boldsymbol{\rm D}^{\star}\right)\right)^{-1}.
 \end{align}
 
 Based on Eq.~(\ref{Problem:P1.5}) and the closed-form optimal solutions of $\boldsymbol{\rm D}^{\star}$ and $\boldsymbol{\rm V}^{\star}$, the problem $\textbf{\textit{P}1}$ can be recast in an equivalent way as follows:
 \begin{subequations}\label{Problem:P2}
 	\begin{alignat}{2}
 	\textbf{\textit{P}1-A:}
 	&\mathop{\min}\limits_{\boldsymbol{\rm W},\boldsymbol{\rm \Gamma}_A} \mbox{Tr}\left(\boldsymbol{\rm W}^H(\boldsymbol{\rm H}_{S}^e)^H\boldsymbol{\rm D}\boldsymbol{\rm V}\boldsymbol{\rm D}^H\boldsymbol{\rm H}_{S}^e\boldsymbol{\rm W}\right)\notag \\
 	&\qquad\qquad+\mbox{Tr}\left(\sigma^2_{\scriptscriptstyle R\hspace{-0.2mm}I\hspace{-0.2mm}S}(\boldsymbol{\rm H}_{S}^n)^H\boldsymbol{\rm D}\boldsymbol{\rm V}\boldsymbol{\rm D}^H\boldsymbol{\rm H}_{S}^n\right)-\mbox{Tr}\left(\boldsymbol{\rm V}\boldsymbol{\rm E}\right)  \notag \\
 	{\rm{s.t.}}:&(\ref{Problem:P1}{\text{a}}), (\ref{Problem:P1}{\text{b}}), ~\text{and}~ (\ref{power_constraint})\notag,
 	\end{alignat}
 \end{subequations}
 where $\boldsymbol{\rm E}=\boldsymbol{\rm D}^H\boldsymbol{\rm H}_{S}^e\boldsymbol{\rm W}+\boldsymbol{\rm W}^H(\boldsymbol{\rm H}_{S}^e)^H\boldsymbol{\rm D}$.
 
 In fact, although $\textbf{\textit{P}1-A}$ eliminates the fractional form of the objective function in $\textbf{\textit{P}1}$, it is still intractable due to the large number of inverses of optimization variables in its objective function and constraints. Therefore, the optimization scheme for conventional active RIS-assisted systems is difficult to be directly applied to our proposed EM-compliant model, and it is imperative to find a suitable optimization scheme.
 
 \section{Proposed Alternating Optimization}
 \label{sec:Alternating_optimization_perfect_CSI}
 
 $\textbf{\textit{P}1}$ is equivalently transformed by the MSE estimation algorithm into the more tractable form $\textbf{\textit{P}1-A}$. To deal with the coupling of variables, we divide the transformed problem into a number of subproblems, where each of the optimization variables is designed alternatively.

 \subsection{Update $\boldsymbol{\rm {W}}$}
 \label{subsec:W}
  
 When the variable $\boldsymbol{\rm {\Gamma}}_A$ is given, substituting the expression for $\boldsymbol{\rm E}$ into $\textbf{\textit{P}1-A}$ and ignoring irrelevant terms, the subproblem with respect to $\boldsymbol{\rm {W}}$ is reconstructed as follows:
 \begin{subequations}\label{Problem:P3}
 	\begin{alignat}{2}
 	\textbf{\textit{P}2:}
 	&\mathop{\min}\limits_{\boldsymbol{\rm W}} \mbox{Tr}\left(\boldsymbol{\rm W}^H\boldsymbol{\rm H}_{1}\boldsymbol{\rm W}\right)-2\mbox{Re}\left\{\mbox{Tr}\left(\boldsymbol{\rm H}_2^H\boldsymbol{\rm W}\right)\right\}  \notag \\
 	{\rm{s.t.}}:&1).\ \mbox{Tr}\left(\boldsymbol{\rm W}\boldsymbol{\rm W}^H\right)\le P_{\max};\\
 	&2).\ \mbox{Tr}\left(\boldsymbol{\rm W}^H\boldsymbol{\rm H}_{3}\boldsymbol{\rm W}\right)\le \bar{P}_{\max}^A,
 	\end{alignat}
 \end{subequations}
  where we define
  \begin{align}
  &\boldsymbol{\rm H}_1 = (\boldsymbol{\rm H}_{S}^e)^H\boldsymbol{\rm D}\boldsymbol{\rm V}\boldsymbol{\rm D}^H\boldsymbol{\rm H}_{S}^e,\notag\\
  &\boldsymbol{\rm H}_2=(\boldsymbol{\rm H}_{S}^e)^H\boldsymbol{\rm D}\boldsymbol{\rm V},~\boldsymbol{\rm H}_3=\left(\boldsymbol{\rm \breve {H}}_{S}^{e}\right)^H\boldsymbol{\rm \breve {H}}_{S}^{e},\notag\\
  &\bar{P}_{\max}^A=P_{\max}^A-\hspace{-1mm}\sigma^2_{\scriptscriptstyle R\hspace{-0.2mm}I\hspace{-0.2mm}S}\mbox{Tr}\hspace{-1mm}\left(\hspace{-1mm}\boldsymbol{\rm \breve {H}}_{S}^{n}\left(\boldsymbol{\rm \breve {H}}_{S}^{n}\right)^H\hspace{-0.8mm}\right) .
  \end{align}
  $\textbf{\textit{P}2}$ is a standard quadratic constraint quadratic programming (QCQP) problem, but still fails to be solved directly with CVX toolkit.

  To facilitate the presentation, we introduce the new definitions $\boldsymbol{\rm {w}}\triangleq\mbox{vec}\left(\boldsymbol{\rm {W}}\right)$, $\boldsymbol{\rm {h}}_2\triangleq\mbox{vec}\left(\boldsymbol{\rm {H}}_2\right)$, $\boldsymbol{\rm \bar {H}}_1\triangleq\boldsymbol{\rm  {I}}_{N_R}\otimes\boldsymbol{\rm {H}}_1$, and $\boldsymbol{\rm \bar {H}}_3\triangleq\boldsymbol{\rm  {I}}_{N_R}\otimes\boldsymbol{\rm {H}}_3$, where $\mbox{vec}\left(\cdot\right)$ and $\otimes$ represent the vectorization and Kronecker product. Thus, $\textbf{\textit{P}2}$ can be equivalently reformulated as
  \begin{subequations}\label{Problem:P3_1}
  	\begin{alignat}{2}
  	\textbf{\textit{P}2-A:}
  	&\mathop{\min}\limits_{\boldsymbol{\rm w}} \boldsymbol{\rm w}^H\boldsymbol{\rm \bar H}_{1}\boldsymbol{\rm w}-2\mbox{Re}\left\{\boldsymbol{\rm h}_2^H\boldsymbol{\rm w}\right\}  \notag \\
  	{\rm{s.t.}}:&1).\ \boldsymbol{\rm w}^H\boldsymbol{\rm w}\le P_{\max};\\
  	&2).\ \boldsymbol{\rm w}^H\boldsymbol{\rm\bar H}_{3}\boldsymbol{\rm w}\le \bar{P}_{\max}^A.
  	\end{alignat}
  \end{subequations}
 $\textbf{\textit{P}2-A}$ is a convex problem that satisfies the criterion of disciplined convex programming (DCP), therefore we can obtain an optimal solution using CVX toolkit.

  \subsection{Update $\boldsymbol{\rm {\Gamma}}_A$}
  \label{subsec:A}
 
  With the variable $\boldsymbol{\rm {W}}$ is fixed, we  investigate the subproblem with respect to $\boldsymbol{\rm {\Gamma}}_A$, which is given as follows:
  \begin{subequations}\label{Problem:P4}
  	\begin{alignat}{2}
  	\textbf{\textit{P}3:}
  	&\mathop{\min}\limits_{\boldsymbol{\rm \Gamma}_A} \mbox{Tr}\left(\boldsymbol{\rm W}^H(\boldsymbol{\rm H}_{S}^e)^H\boldsymbol{\rm D}\boldsymbol{\rm V}\boldsymbol{\rm D}^H\boldsymbol{\rm H}_{S}^e\boldsymbol{\rm W}\right)\notag \\
  	&\qquad\qquad+\mbox{Tr}\left(\sigma^2_{\scriptscriptstyle R\hspace{-0.2mm}I\hspace{-0.2mm}S}(\boldsymbol{\rm H}_{S}^n)^H\boldsymbol{\rm D}\boldsymbol{\rm V}\boldsymbol{\rm D}^H\boldsymbol{\rm H}_{S}^n\right)-\mbox{Tr}\left(\boldsymbol{\rm V}\boldsymbol{\rm E}\right)  \notag \\
  	{\rm{s.t.}}:&\  (\ref{Problem:P1}{\text{b}}) ~\text{and}~ (\ref{power_constraint})\notag.
  	\end{alignat}
  \end{subequations}
  The main challenge in solving $\textbf{\textit{P}3}$ is that the objective function and the constraint of $\textbf{\textit{P}3}$ rely on the inverse of the $\boldsymbol{\rm \Gamma}_A$ matrix, which causes $\boldsymbol{\rm \Gamma}_A$ to be nonconvex in each term of the objective function. For subsequent optimization, we optimize the amplitude matrix of RA and the phase matrix of PS, separately, based on the construction of $\boldsymbol{\rm \Gamma}_A$ in Eq.~(\ref{gamma_3}).

 \subsubsection{Optimize the amplitude matrix $\boldsymbol{\rm \Lambda}_{R\hspace{-0.3mm}A}$}
 
 To deal with the nonlinear dependence of the objective function and the constraint on the reflection coefficients, i.e., the term $\left(\boldsymbol{\rm I}_{M}-\boldsymbol{\rm \Gamma}_{A}\boldsymbol{\rm  S}_{AA}\right)^{-1}\boldsymbol{\rm \Gamma}_{A}$, we propose an BCD method based on Sherman-Morrison inversion, which updates the amplitude $\alpha_{R\hspace{-0.3mm}A,m}$ of the $m$th RA at the $m$th step while keeping all other amplitudes fixed and setting them to the most recently updated values. 
 
 To begin with, we introduce the auxiliary diagonal matrix $\boldsymbol{\rm G}=\boldsymbol{\rm \Gamma}_A^{-1}$, and hence the term $\left(\boldsymbol{\rm I}_{M}-\boldsymbol{\rm \Gamma}_{A}\boldsymbol{\rm  S}_{AA}\right)^{-1}\boldsymbol{\rm \Gamma}_{A}$ can be equivalently rewritten as $\left(\boldsymbol{\rm G}-\boldsymbol{\rm  S}_{AA}\right)^{-1}$. Then, in the $m$th iteration, the inversion $\boldsymbol{\rm G}(m,m)$ of the $m$th reflection coefficient to be optimized is decoupled from the inversions of all other reflection coefficients, which remain fixed. Therefore, we have
 \begin{align}
 \begin{split}
 \left(\boldsymbol{\rm G}-\boldsymbol{\rm  S}_{AA}\right)^{-1}=\left(\boldsymbol{\rm G}_m-\boldsymbol{\rm  S}_{AA}+ g_m\boldsymbol{\rm e}_m\boldsymbol{\rm e}_m^T\right)^{-1},
 \end{split}
 \end{align}
 where $\boldsymbol{\rm G}_m$, $g_m$, and $\boldsymbol{\rm e}_m$ represent the matrix $\boldsymbol{\rm G}$ with $\boldsymbol{\rm G}(m,m)=0$, the $m$th diagonal term of the matrix $\boldsymbol{\rm G}$, and the vector whose terms are all 0, with the exception of the $m$th term being set to 1, respectively. According to Eq.~(\ref{gamma_3}), $g_m$ can be written as 
 \begin{align}\label{alpha_m}
 g_m=\boldsymbol{\rm G}(m,m)=\frac{1}{L_{P\hspace{-0.3mm}S}^{2}e^{j2\theta_m}}\frac{1}{\alpha_{R\hspace{-0.3mm}A,m}}=\frac{1}{q_m}\bar{g}_m,
 \end{align}
 where $q_m=L_{P\hspace{-0.3mm}S}^{2}e^{j2\theta_m}$ and $\bar{g}_m=\alpha_{R\hspace{-0.3mm}A,m}^{-1}$.

 Based on the Sherman-Morrison formula \cite{SE2}, we can derive
 \begin{align}\label{SM}
 \begin{split}
 \left(\boldsymbol{\rm G}-\boldsymbol{\rm  S}_{AA}\right)^{-1}=\boldsymbol{\rm A}_m^{-1}-\frac{\boldsymbol{\rm A}_m^{-1}\boldsymbol{\rm e}_m\boldsymbol{\rm e}_m^{T}\boldsymbol{\rm A}_m^{-1}}{1+g_m\boldsymbol{\rm e}_m^{T}\boldsymbol{\rm A}_m^{-1}\boldsymbol{\rm e}_m}g_m,
 \end{split}
 \end{align}
 where $\boldsymbol{\rm A}_m=\boldsymbol{\rm G}_m-\boldsymbol{\rm  S}_{AA}$. Substituting Eq.~(\ref{SM}) into Eqs.~(\ref{Hs}), (\ref{He}), (\ref{H_A}), and (\ref{H_B}), $\boldsymbol{\rm H}_{S}^{e}$, $\boldsymbol{\rm H}_{S}^{n}$, $\boldsymbol{\rm \breve {H}}_{S}^{e}$, and $\boldsymbol{\rm \breve {H}}_{S}^{n}$ can be reconstructed as follows:
 \begin{align}\label{new_He_Hn}
 \boldsymbol{\rm H}_{S}^{e}=\boldsymbol{\rm B}_{m}+\frac{\boldsymbol{\rm C}_{m}}{d_{m}(\bar{g}_m)},~~~
 \boldsymbol{\rm H}_{S}^{n}=\boldsymbol{\rm \widetilde B}_{m}+\frac{\boldsymbol{\rm \widetilde C}_{m}}{ d_{m}(\bar{g}_m)},\\
 \boldsymbol{\rm \breve H}_{S}^{e}=\boldsymbol{\rm \ddot B}_{m}+\frac{\boldsymbol{\rm \ddot C}_{m}}{d_{m}(\bar{g}_m)},~~~
 \boldsymbol{\rm \breve H}_{S}^{n}=\boldsymbol{\rm \widehat B}_{m}+\frac{\boldsymbol{\rm \widehat C}_{m}}{ d_{m}(\bar{g}_m)},
 \end{align}
 where we define 
 \begin{align}\label{para_suanfa}
 &\boldsymbol{\rm \widetilde A}_m^{-1}\triangleq\frac{\boldsymbol{\rm A}_m^{-1}\boldsymbol{\rm e}_m\boldsymbol{\rm e}_m^{T}\boldsymbol{\rm A}_m^{-1}}{\boldsymbol{\rm e}_m^{T}\boldsymbol{\rm A}_m^{-1}\boldsymbol{\rm e}_m},\notag\\
 &\boldsymbol{\rm B}_{m}\triangleq\boldsymbol{\rm S}_{RT}+\boldsymbol{\rm  S}_{RA}\left(\boldsymbol{\rm  A}_m^{-1}-\boldsymbol{\rm \widetilde A}_m^{-1}\right)\boldsymbol{\rm S}_{AT},\notag\\
 &\boldsymbol{\rm C}_{m}\triangleq\boldsymbol{\rm  S}_{RA}\boldsymbol{\rm \widetilde A}_m^{-1}\boldsymbol{\rm S}_{AT},\notag\\
 &\boldsymbol{\rm \widetilde B}_{m}\triangleq\boldsymbol{\rm  S}_{RA}\left(\boldsymbol{\rm  A}_m^{-1}-\boldsymbol{\rm \widetilde A}_m^{-1}\right),~~\boldsymbol{\rm \widetilde C}_{m}\triangleq\boldsymbol{\rm  S}_{RA}\boldsymbol{\rm \widetilde A}_m^{-1},\notag\\
 &\boldsymbol{\rm  \ddot B}_{m}\triangleq\left(\boldsymbol{\rm  A}_m^{-1}-\boldsymbol{\rm \widetilde A}_m^{-1}\right)\boldsymbol{\rm S}_{AT},~~ \boldsymbol{\rm \ddot C}_{m}\triangleq\boldsymbol{\rm \widetilde A}_m^{-1}\boldsymbol{\rm S}_{AT},\notag\\
 &\boldsymbol{\rm \widehat B}_{m}\triangleq\boldsymbol{\rm  A}_m^{-1}-\boldsymbol{\rm \widetilde A}_m^{-1},~~\boldsymbol{\rm \widehat C}_{m}\triangleq\boldsymbol{\rm \widetilde A}_m^{-1},\notag\\
 & d_{m}(\bar{g}_m)\triangleq 1+a_m\bar{g}_m,~~~~~~ a_m=\boldsymbol{\rm e}_m^{T}\boldsymbol{\rm A}_m^{-1}\boldsymbol{\rm e}_m q_m^{-1}.
 \end{align}
  Substituting Eq.~(\ref{new_He_Hn}) into the objective function and the constraint (\ref{power_constraint}) of $\textbf{\textit{P}3}$, and after a series of algebraic operations, $\textbf{\textit{P}3}$ can be reformulated as follows:
  \begin{subequations}\label{Problem:P4-A}
  	\begin{alignat}{2}
  	\textbf{\textit{P}3-A:}\ 
  	&\mathop{\min}\limits_{\bar{g}_m}~ \frac{c_{1,m}+c_{2,m}\bar{g}_m+c_{2,m}\bar{g}_m^2}{\left(1+a_m^*\bar{g}_m\right)\left(1+a_m\bar{g}_m\right)} \notag \\
  	{\rm{s.t.}}:&1).\  \frac{q_m}{\Gamma_{\max}}\le\bar{g}_m \le q_m, \forall m \in \mathcal{M};\\
  	&2).\ \bar{c}_{1,m}+\bar{c}_{2,m}\bar{g}_m+\bar{c}_{3,m}\bar{g}_m^2\ge 0,
  	\end{alignat}
  \end{subequations}
 where $c_{1,m}$, $c_{2,m}$, $c_{3,m}$, $\bar{c}_{1,m}$, $\bar{c}_{2,m}$, and $\bar{c}_{3,m}$ represent the corresponding constant coefficients independent of $\bar{g}_m$. Obviously, $\textbf{\textit{P}3-A}$ is a single-ratio problem, which can be tackled by employing the Dinkelbach's Transform method \cite{danbi}. Specifically, we introduce the auxiliary variable $\mu_m$ to rewrite  the objective function of $\textbf{\textit{P}3-A}$ as follows:
 \begin{align}
 \begin{split}
 f(\bar{g}_m)=c_{1,m}&+c_{2,m}\bar{g}_m+c_{2,m}\bar{g}_m^2\\
 &-\mu_m\left(1+(a_m+a_m^*)\bar{g}_m+a_ma_m^*\bar{g}_m^2\right),
 \end{split}
 \end{align}
 where $\mu_m$ is iteratively updated at the $(k+1)$th iteration by
 \begin{align}
 \mu_m^{(k+1)}=\frac{c_{1,m}+c_{2,m}\bar{g}_m^{(k)}+c_{2,m}\left(\bar{g}_m^{(k)}\right)^2}{\left(1+a_m^*\bar{g}_m^{(k)}\right)\left(1+a_m\bar{g}_m^{(k)}\right)}.
 \end{align}
 After substituting the objective function, $\textbf{\textit{P}3-A}$ transforms into a constrained polynomial optimization problem. The optimal solution for $\bar{g}_m^{\star}$ can be obtained by using the optimization toolbox in Matlab. Once $\bar{g}_m^{\star}$ is determined for all $m \in \mathcal{M}$, the optimal solution for $\boldsymbol{\rm \Lambda}_{R\hspace{-0.3mm}A}$  can subsequently be recovered.

 \subsubsection{Optimize the phase matrix $\boldsymbol{\rm \Theta}$}
 
 Similarly to $\boldsymbol{\rm \Lambda}_{R\hspace{-0.3mm}A}$, $\boldsymbol{\rm \Theta}$ is also restricted to the nonconvex term $\left(\boldsymbol{\rm I}_{M}-\boldsymbol{\rm \Gamma}_{A}\boldsymbol{\rm  S}_{AA}\right)^{-1}\boldsymbol{\rm \Gamma}_{A}$. To address the subproblem with respect to $\boldsymbol{\rm \Theta}$, we first introduce a small diagonal increment matrix $\boldsymbol{\rm \Delta}$, where $\boldsymbol{\rm \Delta}=\mbox{Diag}\left(\left[\Delta_1,\cdots,\Delta_M\right]^T\right)$  with $\left\vert\Delta_m\right\vert\ll 1$ being the small increment added corresponding to the $m$th phase $\theta_m$. Then, we develop an iteration-based scheme with the matrix $\boldsymbol{\rm \Delta}$, where $\boldsymbol{\rm \Theta}$ is updated by small increments in each iteration. Specifically, the reflection coefficient of the $m$th RE in the $k$th iteration is $\Gamma_{A,m}^{(k)}=L_{P\hspace{-0.3mm}S}^{2}\alpha_{R\hspace{-0.3mm}A,m}e^{j2\theta_m^{(k)}}$. Subsequently, small increment $\Delta_m^{(k)}$ is introduced resulting in the $m$th phase of RE in the $(k+1)$th iteration being changed to $\theta_m^{(k+1)}=\theta_m^{(k)}+\Delta_m^{(k)}$. Therefore, we can obtain the reflection coefficient of the $m$th RE in the $(k+1)$ iteration as follows:
 \begin{align}
 \Gamma_{A,m}^{(k+1)}=L_{P\hspace{-0.3mm}S}^{2}\alpha_{R\hspace{-0.3mm}A,m}e^{j2\left(\theta_m^{(k)}+\Delta_m^{(k)}\right)}.
 \end{align}
 
 Next, utilizing Taylor's series approximation, we have
 \begin{align}
 \Gamma_{A,m}^{(k+1)}\approx \Gamma_{A,m}^{(k)}+j2 L_{P\hspace{-0.3mm}S}^{2}\alpha_{R\hspace{-0.3mm}A,m}e^{j2\left(\theta_m^{(k)}\right)}\Delta_m^{(k)}.
 \end{align}
 Similarly, leveraging Taylor's series approximation, the inverse of $\Gamma_{A,m}^{(k+1)}$ can be written approximately as: 
 \begin{align}
 \left(\Gamma_{A,m}^{(k+1)}\right)^{-1}\hspace{-2.2mm}\approx\hspace{-1mm} \left(\Gamma_{A,m}^{(k)}\right)^{-1}\hspace{-4mm}-j2 L_{P\hspace{-0.3mm}S}^{2}\alpha_{R\hspace{-0.3mm}A,m}\left(\Gamma_{A,m}^{(k)}\right)^{-2}\hspace{-2.5mm}e^{j2\theta_m^{(k)}}\Delta_m^{(k)}.
 \end{align}
 Therefore, extending to the whole matrix of reflection coefficients, we can get
 \begin{align}
 &\left(\boldsymbol{\rm I}_{M}\hspace{-1mm}-\hspace{-1mm}\boldsymbol{\rm \Gamma}_{A}^{(k+1)}\boldsymbol{\rm  S}_{AA}\right)^{-1}\hspace{-2mm}\boldsymbol{\rm \Gamma}_{A}^{(k+1)}=\left(\left(\boldsymbol{\rm \Gamma}_{A}^{(k+1)}\right)^{-1}\hspace{-4mm}-\boldsymbol{\rm  S}_{AA}\right)^{-1}\notag \\
 &\approx\hspace{-1mm}\left( \left(\boldsymbol{\rm \Gamma}_{A}^{(k)}\right)^{-1}\hspace{-4mm}-j2 L_{P\hspace{-0.3mm}S}^{2}\boldsymbol{\rm \Lambda}_{R\hspace{-0.3mm}A}\left(\boldsymbol{\rm \Gamma}_{A}^{(k)}\right)^{-2}\hspace{-2.5mm}e^{j2\boldsymbol{\rm \Theta}^{(k)}}\boldsymbol{\rm \Delta}^{(k)}-\boldsymbol{\rm  S}_{AA}\right)^{-1}\notag\\
 &=\left( \boldsymbol{\rm I}_{M}-\boldsymbol{\rm Y}^{(k)}\boldsymbol{\rm \Delta}^{(k)}\right)^{-1}\left(\boldsymbol{\rm T}^{(k)}\right)^{-1},
 \end{align}
 where we define
 \begin{align}
 &\boldsymbol{\rm T}^{(k)}=\left(\boldsymbol{\rm \Gamma}_{A}^{(k)}\right)^{-1}-\boldsymbol{\rm  S}_{AA},\\
 &\boldsymbol{\rm Y}^{(k)}=j2 L_{P\hspace{-0.3mm}S}^{2}\boldsymbol{\rm \Lambda}_{R\hspace{-0.3mm}A}\left(\boldsymbol{\rm T}^{(k)}\right)^{-1}\left(\boldsymbol{\rm \Gamma}_{A}^{(k)}\right)^{-2}e^{j2\boldsymbol{\rm \Theta}^{(k)}}.
 \end{align}

 However, for the new optimization variable $\boldsymbol{\rm \Delta}^{(k)}$, which is within the inversion of the matrix, the corresponding optimization problem remains nonconvex. To deal with the inversion of the matrix, Neumann series approximation \cite{Schott_Stewart_1999} is introduced, which yields
 \begin{align}\label{Neumann}
 &\left(\boldsymbol{\rm I}_{M}\hspace{-1mm}-\hspace{-1mm}\boldsymbol{\rm \Gamma}_{A}^{(k+1)}\boldsymbol{\rm  S}_{AA}\right)^{-1}\hspace{-2mm}\boldsymbol{\rm \Gamma}_{A}^{(k+1)}\approx\hspace{-1mm}\left(\boldsymbol{\rm T}^{(k)}\right)^{-1}\hspace{-3mm}+\boldsymbol{\rm Y}^{(k)}\boldsymbol{\rm \Delta}^{(k)}\hspace{-1mm}\left(\boldsymbol{\rm T}^{(k)}\right)^{-1}\hspace{-3mm},
 \end{align}
 which is accurate if $\left\Vert\boldsymbol{\rm Y}^{(k)}\boldsymbol{\rm \Delta}^{(k)}\right\Vert\ll1$.
 Thus, substituting Eq.~(\ref{Neumann}) into Eq.~(\ref{Hs}) and (\ref{He}), $\boldsymbol{\rm H}_{S}^{e}$ and $\boldsymbol{\rm H}_{S}^{n}$ can be reconstructed as follows:
 \begin{align}\label{2_He}
 &\boldsymbol{\rm H}_{S}^{e}=\boldsymbol{\rm S}_{1}^{(k)}+\boldsymbol{\rm S}_{2}^{(k)}\boldsymbol{\rm \Delta}^{(k)}\boldsymbol{\rm S}_{3}^{(k)},\\
 &\boldsymbol{\rm H}_{S}^{n}=\boldsymbol{\rm \widetilde S}_{1}^{(k)}+\boldsymbol{\rm S}_{2}^{(k)}\boldsymbol{\rm \Delta}^{(k)}\boldsymbol{\rm \widetilde S}_{3}^{(k)},\label{2_Hn}
 \end{align}
 where the parameters are defined as follows:
 \begin{align}
 &\boldsymbol{\rm S}_{1}^{(k)}\triangleq\boldsymbol{\rm S}_{RT}+\boldsymbol{\rm  S}_{RA}\left(\boldsymbol{\rm T}^{(k)}\right)^{-1}\boldsymbol{\rm S}_{AT},\notag\\
 &\boldsymbol{\rm S}_{2}^{(k)}\triangleq\boldsymbol{\rm  S}_{RA}\boldsymbol{\rm Y}^{(k)},~~~~~~~~~~~~\boldsymbol{\rm S}_{3}^{(k)}\triangleq \left(\boldsymbol{\rm T}^{(k)}\right)^{-1}\boldsymbol{\rm S}_{AT},\notag\\
 &\boldsymbol{\rm \widetilde S}_{1}^{(k)}\triangleq\boldsymbol{\rm  S}_{RA}\left(\boldsymbol{\rm T}^{(k)}\right)^{-1},~~~~~\boldsymbol{\rm \widetilde S}_{3}^{(k)}\triangleq \left(\boldsymbol{\rm T}^{(k)}\right)^{-1}.
 \end{align}
 
  Then, let $\boldsymbol{\rm \bar \Delta}=\left[\Delta_1,\cdots,\Delta_M\right]^T.$ 
  Substituting Eqs.~(\ref{2_He}) and (\ref{2_Hn}) into the objective function of $\textbf{\textit{P}3}$, and after a series of algebraic operations, the subproblem with respect to $\boldsymbol{\rm \bar\Delta}^{(k)}$ in the $(k+1)$th iteration can be rewritten as follows:
  \begin{subequations}\label{Problem:P4-B}
 	\begin{alignat}{2}
 	\textbf{\textit{P}3-B:}
 	&\mathop{\min}\limits_{\boldsymbol{\rm \bar\Delta}^{(k)}} \left(\boldsymbol{\rm \bar\Delta}^{(k)}\right)^{H}\boldsymbol{\rm {\Upsilon}}^{(k)}\boldsymbol{\rm \bar\Delta}^{(k)}+2\mbox{Re}\left\{\boldsymbol{\rm \bar\Delta}^{(k)}\left(\boldsymbol{\rm \zeta}^{(k)}\right)^*\right\}\notag \\
 	{\rm{s.t.}}:&\  \left\Vert\boldsymbol{\rm Y}^{(k)}\boldsymbol{\rm \Delta}^{(k)}\right\Vert\ll1,
 	\end{alignat}
 \end{subequations}
 where we define
 \begin{align}
 &\boldsymbol{\rm {\Upsilon}}^{(k)}=\left(\hspace{-1mm}\left(\boldsymbol{\rm S}_{2}^{(k)}\hspace{-1mm}\right)^H\hspace{-2mm}\boldsymbol{\rm D}\boldsymbol{\rm V}\boldsymbol{\rm D}^H\boldsymbol{\rm S}_{2}^{(k)}\hspace{-1mm}\right)\hspace{-1mm}\odot\hspace{-1mm}\left(\hspace{-1mm}\boldsymbol{\rm S}_{3}^{(k)}\boldsymbol{\rm W}\boldsymbol{\rm W}^H\hspace{-1mm}\left(\boldsymbol{\rm S}_{3}^{(k)}\hspace{-1mm}\right)^H\hspace{-1mm}\right)^T\notag\\
 &+ \left(\hspace{-1mm}\sigma^2_{\scriptscriptstyle R\hspace{-0.2mm}I\hspace{-0.2mm}S}\left(\boldsymbol{\rm\widetilde S}_{2}^{(k)}\hspace{-1mm}\right)^H\hspace{-2mm}\boldsymbol{\rm D}\boldsymbol{\rm V}\boldsymbol{\rm D}^H\boldsymbol{\rm\widetilde S}_{2}^{(k)}\hspace{-1mm}\right)\hspace{-1mm}\odot\hspace{-1mm}\left(\hspace{-1mm}\sigma^2_{\scriptscriptstyle R\hspace{-0.2mm}I\hspace{-0.2mm}S}\boldsymbol{\rm\widetilde S}_{3}^{(k)}\hspace{-1mm}\left(\boldsymbol{\rm\widetilde S}_{3}^{(k)}\hspace{-1mm}\right)^H\hspace{-1mm}\right)^T\hspace{-1mm},\\
 &\boldsymbol{\rm \zeta}^{(k)}=\mbox{Diag}\left(\hspace{-1mm}\boldsymbol{\rm S}_{3}^{(k)}\boldsymbol{\rm W}\boldsymbol{\rm W}^H\left(\boldsymbol{\rm S}_{1}^{(k)}\hspace{-1mm}\right)^H\hspace{-2mm}\boldsymbol{\rm D}\boldsymbol{\rm V}\boldsymbol{\rm D}^H\boldsymbol{\rm S}_{2}^{(k)}\hspace{-1mm}\right)\notag\\
 &\qquad\quad+ \mbox{Diag}\left(\hspace{-1mm}\sigma^2_{\scriptscriptstyle R\hspace{-0.2mm}I\hspace{-0.2mm}S}\boldsymbol{\rm\widetilde S}_{3}^{(k)}\left(\boldsymbol{\rm\widetilde S}_{1}^{(k)}\hspace{-1mm}\right)^H\hspace{-2mm}\boldsymbol{\rm D}\boldsymbol{\rm V}\boldsymbol{\rm D}^H\boldsymbol{\rm\widetilde S}_{2}^{(k)}\hspace{-1mm}\right)\notag\\
 &\qquad\quad- \mbox{Diag}\left(\hspace{-1mm}\boldsymbol{\rm S}_{3}^{(k)}\boldsymbol{\rm W}\boldsymbol{\rm V}\boldsymbol{\rm D}^H\boldsymbol{\rm S}_{2}^{(k)}\hspace{-1mm}\right).
 \end{align}
 Further, we utilize the properties of $\boldsymbol{\rm \bar\Delta}^{(k)}$ to simplify $\textbf{\textit{P}3-B}$. To begin with, since $\left\vert\Delta_m^{(k)}\right\vert\ll 1$, the quadratic term about $\boldsymbol{\rm \bar\Delta}^{(k)}$ in the objective function can be ignored. In addition, for the constraint (\ref{Problem:P4-B}{a}), we can define the small increment ${ \Delta}_0=\left\Vert\boldsymbol{\rm Y}^{(k)}\boldsymbol{\rm \Delta}^{(k)}\right\Vert\ll 1$, which is added to all elements of $\boldsymbol{\rm \Delta}^{(k)}$ to ensure that the constraint is satisfied. Thereby, the constraint (\ref{Problem:P4-B}{a}) can be equivalently rewritten as
 \begin{align}\label{constraint1}
 \left\vert\Delta_m^{(k)}\right\vert\le\Delta_0\left\Vert\boldsymbol{\rm Y}^{(k)}\right\Vert^{-1}, \forall m \in\mathcal{M}.
 \end{align}

 Finally, the $\textbf{\textit{P}3-B}$ breaks down to $M$ independent small problems with the $m$th small problem written as follows:
 \begin{subequations}\label{Problem:P4-Bm}
 	\begin{alignat}{2}
 	\textbf{\textit{P}3-B(m):}
 	&\mathop{\min}\limits_{\Delta^{(k)}_m}~~2\mbox{Re}\left\{\left( \zeta^{(k)}_m\right)^*\right\}\Delta^{(k)}_m\notag \\
 	{\rm{s.t.}}:&\  \left\vert\Delta_m^{(k)}\right\vert\le\Delta_0\left\Vert\boldsymbol{\rm Y}^{(k)}\right\Vert^{-1},
 	\end{alignat}
 \end{subequations}
 where $\zeta^{(k)}_m$ denotes the $m$th element of $\boldsymbol{\rm \zeta}^{(k)}$. Thus, we can obtain the optimal solution of $\Delta^{(k)}_m$ as follows:
 \begin{align}\label{solve_theta}
 \left\{
 \begin{aligned}
 &{\Delta^{(k)}_m}^{\star}=\Delta_0\left\Vert\boldsymbol{\rm Y}^{(k)}\right\Vert^{-1}, ~\text{if}~\mbox{Re}\left\{\left( \zeta^{(k)}_m\right)^*\right\}<0;\\
 &{\Delta^{(k)}_m}^{\star}=-\Delta_0\left\Vert\boldsymbol{\rm Y}^{(k)}\right\Vert^{-1},~\text{if}~\mbox{Re}\left\{\left( \zeta^{(k)}_m\right)^*\right\}\ge 0.
 \end{aligned}
 \right.
 \end{align}
 Therefore, the $m$th phase of RE in the $(k+1)$th iteration is computed as $\theta_m^{(k+1)}=\theta_m^{(k)}+{\Delta^{(k)}_m}^{\star}$, and then $\Gamma_{A,m}^{(k+1)}$ is deduced from $\theta_m^{(k+1)}$. The process does not terminate until convergence.
 
 \begin{algorithm}[t]
 	\caption{SMaN-based AO algorithm for Problem \textbf{\textit{P}1}.} \label{alg:Framwork_all}
 	\begin{algorithmic}[1]
 		\STATE Initialize $\boldsymbol{\rm {W}}^{(0)}$,$\boldsymbol{\rm \Lambda}_{R\hspace{-0.3mm}A}^{(0)}$,$\boldsymbol{\rm {\Theta}}^{(0)}$, the accuracy $\eta_{\varsigma}=10^{-3}$, the number of iterations $k=0$, and the maximum number of iterations $k_{\max}$;
 		\REPEAT
 		\STATE Compute the objective function $V_{OF} (k)$ of $\textbf{\textit{P}1}$;
 		\STATE Update $\boldsymbol{\rm {D}}^{(k+1)}$ and $\boldsymbol{\rm {V}}^{(k+1)}$ by (\ref{para_D}) and (\ref{para_V}), respectively;
 		\STATE Tackle the subproblem $\textbf{\textit{P}2-A}$ and obtain the optimal solution $\boldsymbol{\rm {W}}^{(k+1)}$;
 		\FOR{$m=1$ to $M$}
 		\STATE Compute $\bar{g}_m^{\star}$ by solving $\textbf{\textit{P}3-A}$;
 		\STATE Update $\boldsymbol{\rm \Lambda}_{R\hspace{-0.3mm}A}^{\star}(m,m)$ by computing Eq.~(\ref{alpha_m});
 		\ENDFOR
 		\STATE Update $\boldsymbol{\rm \Lambda}_{R\hspace{-0.3mm}A}^{(k+1)}=\boldsymbol{\rm \Lambda}_{R\hspace{-0.3mm}A}^{\star}$;
 		\STATE Compute the subproblem $\textbf{\textit{P}3-B(m)}$ and obtain the optimal solution $\boldsymbol{\rm {\Delta}}^{(k)}$ by Eq.~(\ref{solve_theta});
 		\STATE Update $\boldsymbol{\rm {\Theta}}^{(k+1)}=\boldsymbol{\rm {\Theta}}^{(k)}+\boldsymbol{\rm {\Delta}}^{(k)}$;
 		\STATE Recover $\boldsymbol{\rm \Gamma}_A^{(k+1)}=L_{P\hspace{-0.3mm}S}^{2}\boldsymbol{\rm \Lambda}_{R\hspace{-0.3mm}A}^{(k+1)}\exp\left(j2\boldsymbol{\rm \Theta}^{(k+1)}\right)$;
 		\STATE update $k=k+1$;
 		\UNTIL $\left\vert V_{OF} (k+1)-V_{OF} (k)\right\vert<\eta_{\varsigma}$ or $k = k_{\max}$.
 	\end{algorithmic}
 \end{algorithm}
 
 The complete AO algorithm based on Sherman-Morrison inversion and Neumann series (abbreviated as SMaN-based AO algorithm) for solving problem $\textbf{\textit{P}1}$ is summarized in Algorithm~\ref{alg:Framwork_all}. Then, we investigate the complexity of the Algorithm~\ref{alg:Framwork_all}. In each iteration, the main complexity is embodied in the optimization of $\boldsymbol{\rm {W}}$, $\boldsymbol{\rm \Lambda}_{R\hspace{-0.3mm}A}$ and $\boldsymbol{\rm {\Theta}}$. The complexity of optimizing $\boldsymbol{\rm W}$ is governed by the computation of the matrix inversion in the computation of $\boldsymbol{\rm W}^{\star}$ as well as the solution process of CVX, which is given as $\mathcal{O}_1=\mathcal{O}\left(N_T^3+\left(N_TN_R\right)^{3.5}\right)$. The complexity of optimizing $\boldsymbol{\rm \Lambda}_{R\hspace{-0.3mm}A}$ is determined by the computation of the parameters in Eq.~(\ref{para_suanfa}), which has a total complexity of $\mathcal{O}_2=\mathcal{O}\left(M\left(M^3+M^2N_R\right)\right)$. Besides, the complexity of optimizing $\boldsymbol{\rm \Theta}$ comes from computing the inversion of $\boldsymbol{\rm T}^{(k)}$, whose complexity is $\mathcal{O}_3=\mathcal{O}\left(M^3\right)$. As a result, the complexity of the  overall Algorithm~\ref{alg:Framwork_all} is given as $\mathcal{O}\left(k_{\max}\left(\mathcal{O}_1+\mathcal{O}_2+\mathcal{O}_3\right)\right)$.

\section{Simulation Results and Analysis}
\label{sec:simulation}

\begin{figure}[t]
 	\centering
 	\includegraphics[scale=0.042]{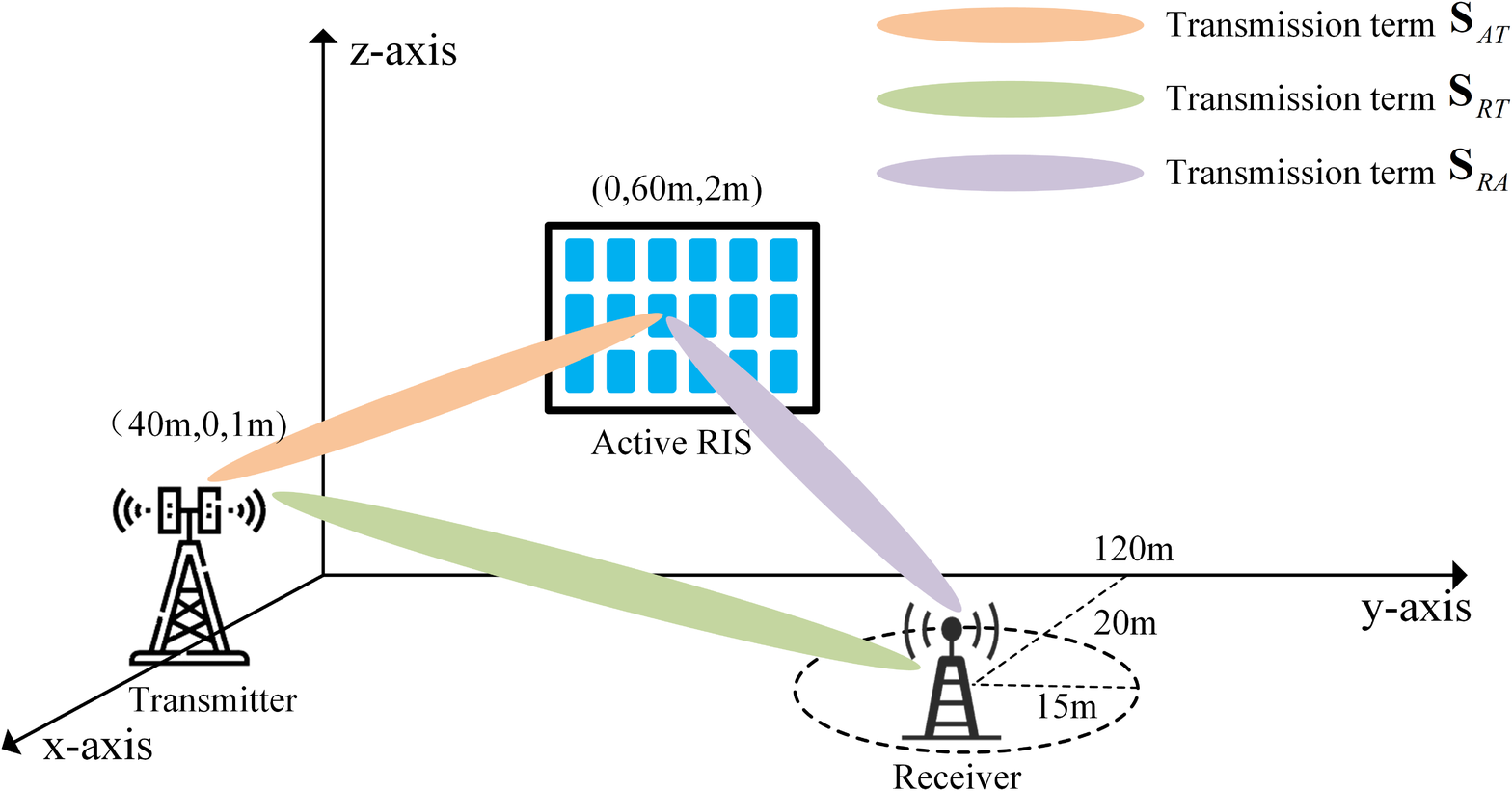}
 	\vspace{-10pt}
 	\caption{Plane diagram of simulated active RIS-assisted wireless communication system.}
 	\vspace{-13pt}
 	\label{fig:system_model}
 \end{figure}

 In this section, we conduct the simulations to evaluate the performance of EM-compliant model with MC for active RIS-assisted communication at $2.4$ GHz. Unless otherwise noted, parameters are set throughout the simulation. In the system, the transmitter equipped with $N_T=4$ antennas communicates with the receiver equipped with $N_R = 2$ with the assistance of the active RIS. To facilitate the elaboration of the location of the devices in the system, we establish a coordinate system as in Fig.~\ref{fig:system_model}, where the coordinates of transmitter and the center of the uniform planar array (UPA) at active RIS are (40m, 0, 1m) and (0, 60m, 2m), respectively. The receiver is located on circles of radius 15 centered at (20m, 120m, 1m). Besides, the RE spacing at the active RIS is set as $d_{s}=\lambda_c/4$, where $\lambda_c$ denotes the wavelength. 
 
\begin{table}[t]
	\renewcommand{\arraystretch}{1.5}
	\caption{The value of $S$-parameter of the active RIS at 2.4 GHz.}
	\label{table_1}
	\centering
	\begin{tabular}{|c|c|c|c|c|}
		\hline
		\diagbox{\quad}{$d_{s}$} & $\lambda_c/2$ &$\lambda_c/3$ & $\lambda_c/4$ & $\lambda_c/6$\\
		\hline
		$S(1,1)$ (dB) & -25.15&-24.26&-20.81&-16.23\\
		\hline
		$S(2,1)$ (dB) & -19.11&-18.25&-13.62&-12.43\\
		\hline
	\end{tabular}
\end{table}

 \begin{figure}[t]
 	\vspace{-15pt}
 	\centering
 	\includegraphics[scale=0.50]{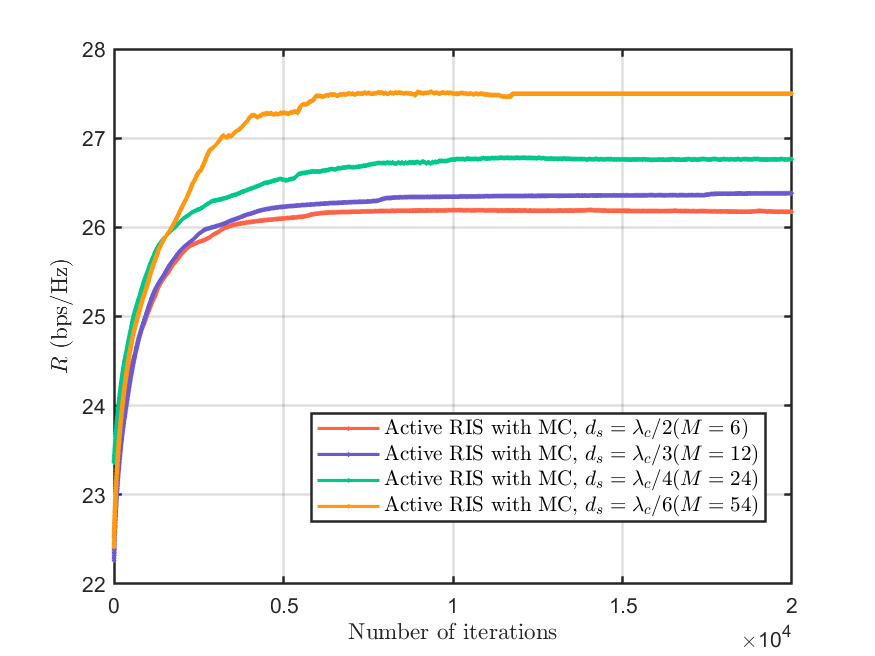}
 	\vspace{-10pt}
 	\caption{Convergence with fixed active RIS size.}
 	\vspace{-13pt}
 	\label{fig:ite_EE}
 \end{figure}

\begin{figure}[t]
	\centering
	\includegraphics[scale=0.50]{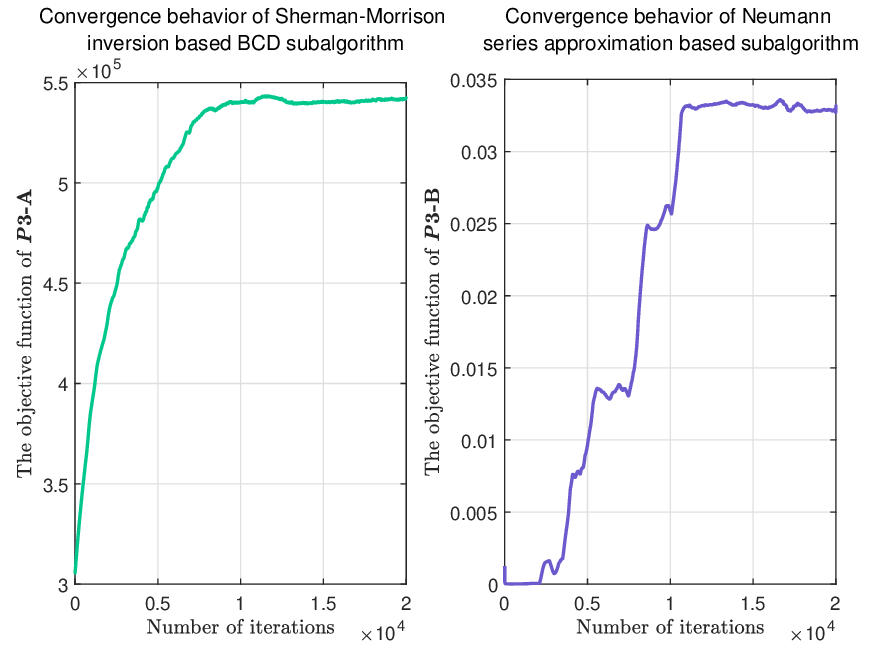}
	\vspace{-10pt}
	\caption{(a) Convergence behavior of Sherman-Morrison inversion based BCD subalgorithm. (b) Convergence behavior of Neumann series approximation based subalgorithm.}
	\vspace{-13pt}
	\label{fig:ite_1}
\end{figure}
 
 We construct the practical active RIS that adheres to the specifications in Section~\ref{subsec:EM_Active_RIS} and then extract its key parameters\footnote{The designed active RIS in \cite{Active_RIS_cankao} serves as a representative example for constructing the active RIS in this paper.}. In particular, its $S$-parameter matrix containing MC and mismatch effects, ie., $\boldsymbol{\rm {S}}_{AA}$, is applied to the optimization problem. With the analytical models presented in \cite{Renzo_2021} and \cite{direnzo2023modeling} as well as the full-wave simulator, the $S$-parameter matrix can be simulated and computed as shown in Table~\ref{table_1}. Specifically, $S(1,1)$ and $S(2,1)$ in the table can characterize the matching and MC terms , i.e., diagonal and non-diagonal elements, of $\boldsymbol{\rm {S}}_{AA}$, respectively. In addition, we set the maximum amplification gain of the reflection coefficient as $\Gamma_{\max}^2=30$ dB. For brevity, we assume that the noise power is $\sigma^2=-100$ dBm. The noise power introduced by the active RIS is set to $\sigma^2_{\scriptscriptstyle R\hspace{-0.2mm}I\hspace{-0.2mm}S}= -100$ dBm.
 
 Transmission terms $\boldsymbol{\rm {S}}_{RT}$, $\boldsymbol{\rm {S}}_{AT}$, and $\boldsymbol{\rm {S}}_{RA}$ include random transmission terms and the main hardware characteristics (MC and mismatch) of the corresponding ports. In this paper, we assume that the transmitter and receiver are in an impedance-matched configuration, thus we can model the transmission terms by utilizing the Rician channel model \cite{Shen_2022}, as follows:
 \begin{align}
 \boldsymbol{\rm {S}}_{\bar\chi}=\sqrt{PL_{\bar\chi}}\left(\sqrt{\frac{\kappa}{\kappa+1}}\boldsymbol{\rm {S}}_{\bar\chi}^{\rm LOS}+\sqrt{\frac{1}{\kappa+1}}\boldsymbol{\rm {S}}_{\bar\chi}^{\rm NLOS}\right),
 \end{align}
 where $PL_{\bar\chi}$, $\kappa$, $\boldsymbol{\rm {S}}_{\bar\chi}^{\rm LOS}$, and $\boldsymbol{\rm {S}}_{\bar\chi}^{\rm NLOS}$ denote the path loss, Rician factor, the LOS component, and the NLOS component of the transmission term $\boldsymbol{\rm {S}}_{\bar\chi}$, respectively, where $\bar\chi\in\left\{RT,AT,RA\right\}$. The NLOS component $\boldsymbol{\rm {S}}_{\bar\chi}^{\rm NLOS}$ follows the Rayleigh fading model. The LOS component $\boldsymbol{\rm {S}}_{\bar\chi}^{\rm LOS}$ represents the array response, which can be written as $\boldsymbol{\rm {\Lambda}}_{\widetilde{N}}(\varphi)=\left[1,e^{j2\pi\frac{d}{\lambda_c}\sin(\varphi)},\cdots,e^{j2\pi\frac{d}{\lambda_c}(\widetilde{N}-1)\sin(\varphi)}\right]^H$, where $\widetilde{N}=\left\{N_T,M\right\}$ and $\varphi$ indicates the Angle of Arrival (AoA) or the Angle of Departure (AoD). Thus, we have $\boldsymbol{\rm {S}}_{RT}^{\rm LOS}=\boldsymbol{\rm {\Lambda}}_{N_T}(\iota_T)$, $\boldsymbol{\rm {S}}_{AT}^{\rm LOS}=\boldsymbol{\rm {\Lambda}}_{M}(\varphi_T)\boldsymbol{\rm {\Lambda}}_{N_T}(\iota_A)^H$, and $\boldsymbol{\rm {S}}_{RA}^{\rm LOS}=\boldsymbol{\rm {\Lambda}}_{M}(\iota_R)$, where $\iota_T,\iota_T,\iota_T$ are the AoD and $\varphi_T$ is the AoA. The path loss of the transmission term in dB is defined as $PL_{\bar\chi} = \beta_0 +10\beta_{\bar\chi} \mbox{log}_{10}(d_{\bar\chi}/d_0)$, where $\beta_0$, $\beta_{\bar\chi}$ and $d_{\bar\chi}$ are the path loss at the reference distance $d_0=1$ m, the path loss exponent, and the distance between corresponding devices. According to \cite{Energy-Effic}, we employ $\beta_{RT} = 2.45$, $\beta_{RA}= 2.2$, and $\beta_{AT} = 2.2$ to denote the path loss exponents of transmission terms $\boldsymbol{\rm {S}}_{RT}$, $\boldsymbol{\rm {S}}_{RA}$, and $\boldsymbol{\rm {S}}_{AT}$, respectively. All simulation curves are derived from the average of 100 random tests.

 For the purpose of comparative analysis, we compare our proposed EM-compliant model for active RIS with MC (shortened as active RIS with MC) with other RIS models, including active RIS without MC \cite{Active_RIS1}, active RIS with MC unawareness, and passive RIS with/without MC \cite{S_parameter}. The active RIS without MC represents conventional active RIS model, where the MC effect is assumed to be absent at the active RIS. The active RIS with MC unawareness refers to the scenario in which the optimal values derived from the conventional active RIS model are implanted into the practical setup of the EM-compliant model with MC. The passive RIS with/without MC refers to conventional passive RIS model in the presence/absence of the MC effect. To ensure fairness, the transmit power budget in the system associated with the passive RIS is defined as the sum of the power budgets of the active RIS and the transmitter.

 \begin{figure}[t]
 	\vspace{-15pt}
 	\centering
 	\includegraphics[scale=0.50]{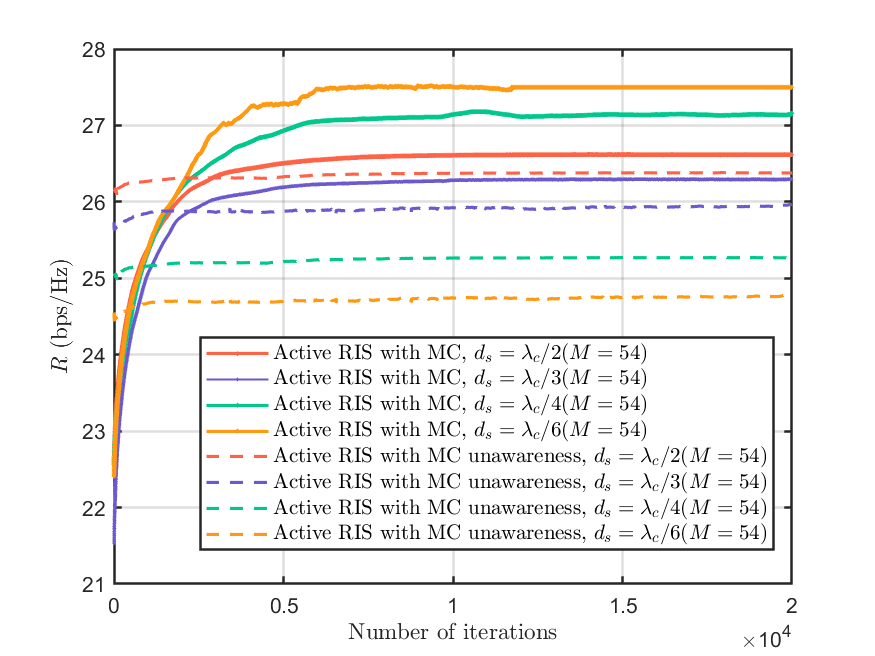}
 	\vspace{-10pt}
 	\caption{Convergence with fixed number of REs}
 	\vspace{-13pt}
 	\label{fig:ite_R_1}
 \end{figure}


  Figure~\ref{fig:ite_EE} illustrates the convergence behavior of the Algorithm~\ref{alg:Framwork_all} for different values of $M$ and $d_s$. As the number of iterations increases, the achievable rate $R$ of the system increases significantly and then gradually converges, which validates our derived theorem and designed algorithm. In addition, it can be seen that the convergence speed of Algorithm~\ref{alg:Framwork_all} depends on $M$ and $d_s$. For the fixed active RIS size, smaller spacing distances (a greater number of REs) make the algorithm converge slower \footnote{With the size of RIS fixed, the number of REs varies with the spacing distance among REs. In addition, the level of the MC effect is directly dependent on the RE spacing. As the spacing distance is closer, neighboring REs will couple more waves.}. Furthermore, the achievable rate grows as the number of REs increases. This is because the growth in the number of REs effectively compensates for the reduction in $R$ due to the increased MC. Figure~\ref{fig:ite_1} further demonstrates the convergence of our proposed SMaN-based AO algorithm. The proposed SMaN-based algorithm consists of Sherman-Morrison inversion based BCD subalgorithm and Neumann series approximation based subalgorithm, both of which are required to converge to a steady state in order to prove the convergence of Algorithm~\ref{alg:Framwork_all}. As shown in Fig.~\ref{fig:ite_1}, we can observe that the objective function values of the two subproblems ($\textbf{\textit{P}3-A}$ and $\textbf{\textit{P}3-B}$) are gradually stabilized after iterations, verifying the convergence of their corresponding subalgorithms.

  The convergence behavior of Algorithm~\ref{alg:Framwork_all} for the same number of REs at the active RIS is further shown in Fig.~\ref{fig:ite_R_1}, where the active RIS with MC unawareness scheme is utilized for comparison with our proposed active RIS with MC scheme. In the active RIS with MC scheme, the achievable rate features a non-monotonic behavior of $d_s$, due to the combined influence of the size of the active RIS and the MC effect. In contrast, in the active RIS with MC unawareness scheme, a decrease in $d_s$ entails a decrease in the achievable rate. It is because as $d_s$ decreases, the enhanced MC effect restricts the system performance. Furthermore, we can observe that the performance gap between the developed active RIS with MC scheme and the active RIS with MC unawareness scheme grows as the RE spacing at the active RIS decreases. This is because the MC effect becomes more significant when the RE spacing is reduced, whereas the active RIS with MC scheme can effectively eliminate the performance degradation caused by MC effect.

  \begin{figure}[t]
  	\vspace{-15pt}
  	\centering
  	\includegraphics[scale=0.50]{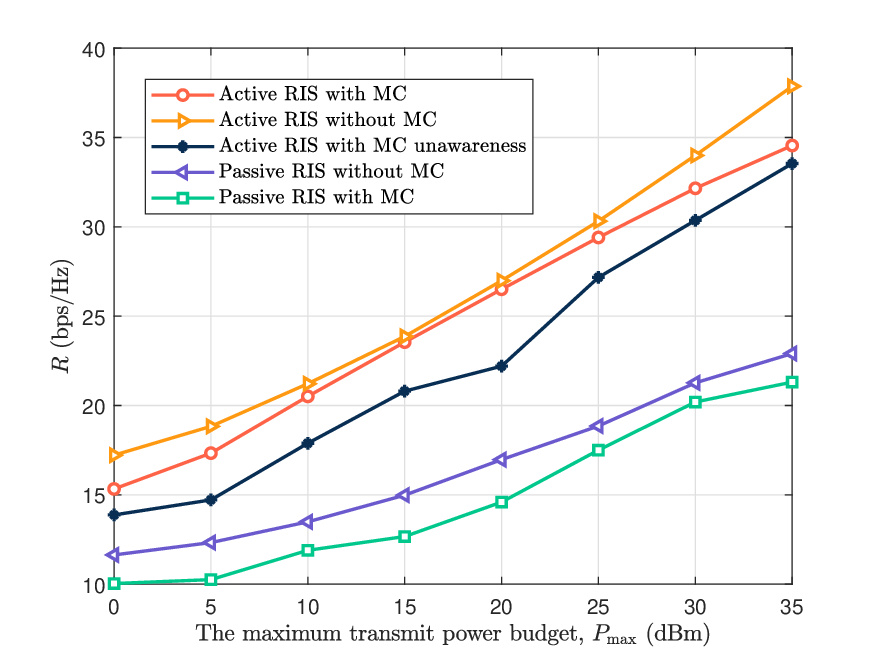}
  	\vspace{-10pt}
  	\caption{The achievable rate $R$ versus the maximum transmit power budget $P_{\max}$, where $d_s=\lambda_c/4$ and $P_{\max}^A=30$ dB.}
  	\vspace{-13pt}
  	\label{fig:P_BS_EE}
  \end{figure}
 
  Figure~\ref{fig:P_BS_EE} shows the achievable rate $R$ versus the maximum transmit power budget $P_{\max}$ for different schemes. The achievable rate increases with $P_{\max}$ under different schemes, but the trend of increase varies. Compared to passive RIS, the effective amplification of the reflected signal by active RIS creates a significant increase in $R$ as $P_{\max}$ grows. The performance of the active RIS with MC scheme approximates the performance of the ideal active RIS without MC scheme, which verifies that our proposed EM-compliant active RIS model still guarantees sufficient performance improvement while taking into account the EM characteristics. Besides, the optimization bias from ignoring the MC effect leads to the performance gap between the active RIS with MC scheme and the active RIS with MC unawareness scheme.
  
 \begin{figure}[t]
 	\centering
 	\includegraphics[scale=0.50]{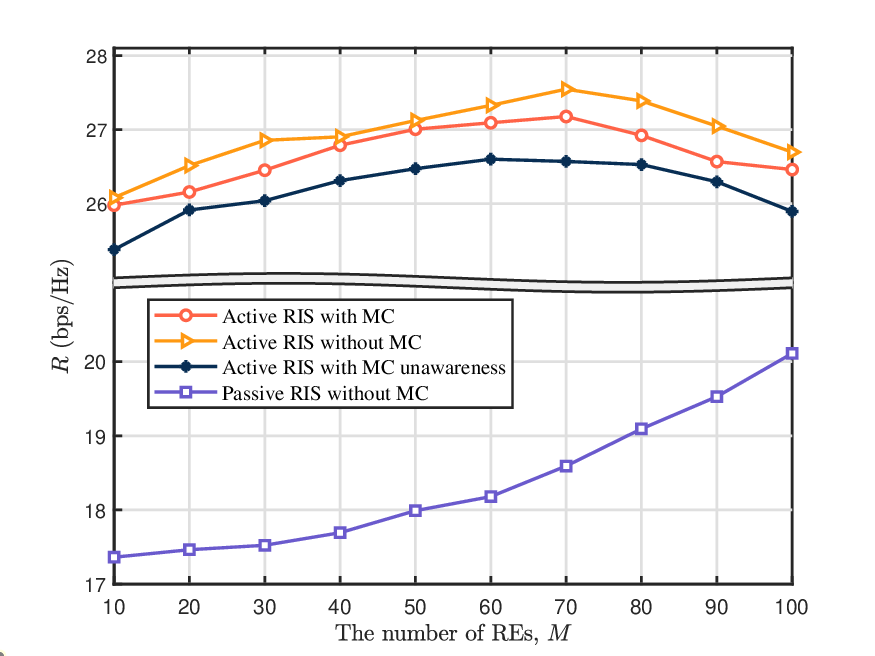}
 	\vspace{-10pt}
 	\caption{The achievable rate $R$ versus the number $M$ of REs at the RIS, where $d_s=\lambda_c/4$, $P_{\max}=20$ dB, and $P_{\max}^A=20$ dB.}
 	\vspace{-13pt}
 	\label{fig:M_R}
 \end{figure}
  
  Figure~\ref{fig:M_R} depicts the achievable rate $R$ versus the number $M$ of REs at the RIS, where the spacing of neighboring REs is kept constant while expanding the size of the EM-compliant active RIS. The performance differences between the different schemes are the same as in Fig.~\ref{fig:P_BS_EE}. As the number of RE increases, the $R$ of the active RIS first rises to a peak and then slowly decreases. As the number $M$ of REs increases, the $R$ of the active RIS first rises to a peak and then slowly decreases. This is because the amplification power budget $P_{\max}^A$ initially caters for a smaller number of REs to adequately amplify the reflected signal (i.e., $\left\vert\Gamma_{A,m}\right\vert = \Gamma_{\max}, \forall m \in \mathcal{M}$), thus $R$ benefits more from the initial increase in $M$ up to the peak. Then, as M continues to increase, $P_{\max}^A$ struggles to support adequate amplification of signals by all the REs. As a result, due to the limitations of the amplification power budget, the inadequate amplification of signals by more REs beyond the peak point, resulting in a decrease in $R$. This trend reflects the trade-off between the amplification power budget and the maximum amplitude of the reflection coefficient at the active RIS. Therefore, the settings of $P_{\max}^A$ and $\Gamma_{\max}$ need to be carefully considered when designing active RIS.

  \begin{figure}[t]
  	\vspace{-15pt}
  	\centering
  	\includegraphics[scale=0.50]{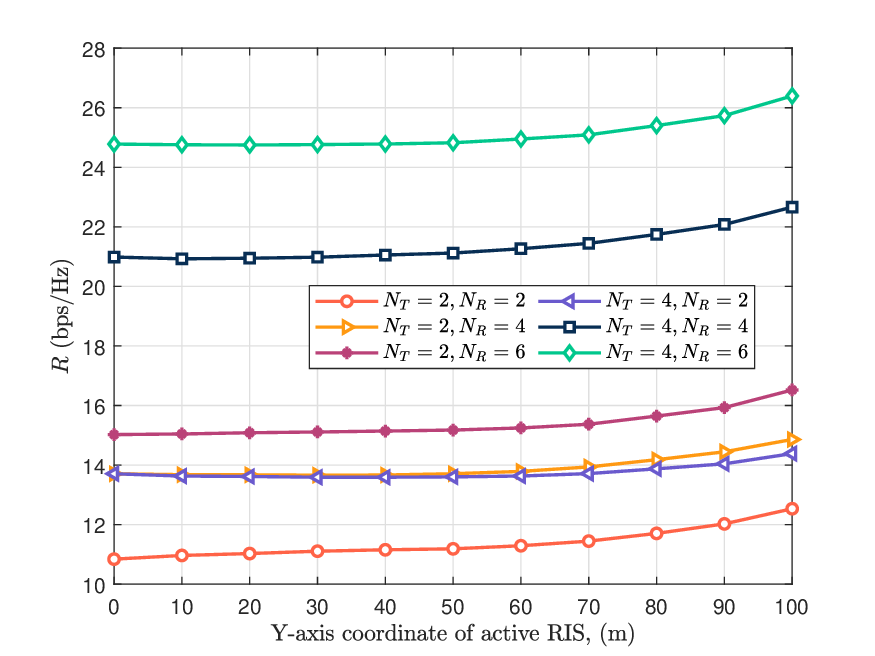}
  	\vspace{-10pt}
  	\caption{The achievable rate $R$ versus the Y-axis coordinate of active RIS, where $d_s=\lambda_c/4$, $P_{\max}=5$ dB, and $P_{\max}^A=5$ dB.}
  	\vspace{-13pt}
  	\label{fig:L_EE}
  	\end{figure}

 Figure~\ref{fig:L_EE} compares the achievable rate $R$ of the active RIS-assisted system with the position of the EM-compliant active RIS, which varies with the Y-axis coordinate of active RIS. It can be seen that the achievable rate increases as the Y-axis coordinate of the active RIS increases. This is due to the fact that as the EM-compliant active RIS gets closer to the receiver, the power of the impinging signal at the active RIS becomes weaker, thus allowing it to benefit more from the amplification of the active RIS, i.e., the active RIS can provide more effective amplification. This result reveals that positioning the active RIS near the receiver yields a more significant performance gain. In addition, the effect of different numbers of transmitter and receiver antennas on the performance of the system is also shown in Fig.~\ref{fig:L_EE}. $R$ increases as the number of transmitter and receiver antennas grows. This is because the increase in the number of antennas enhances the diversity gain.
 
 \begin{figure}[t]
 	\centering
 	\includegraphics[scale=0.50]{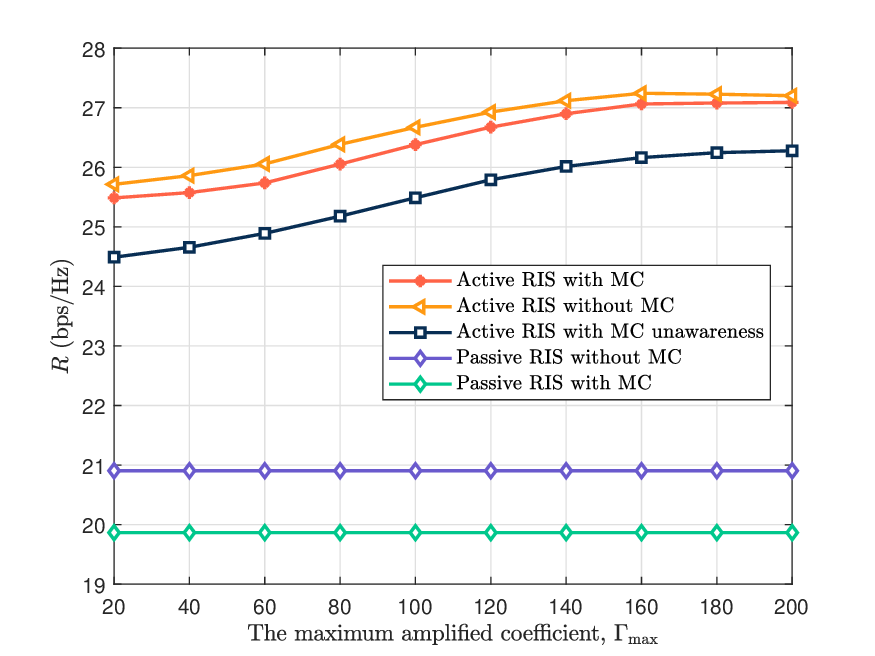}
 	\vspace{-10pt}
 	\caption{The achievable rate $R$ versus the maximum amplified coefficient, $\Gamma_{\max}$, where $d_s=\lambda_c/4$, $P_{\max}=20$ dB, and $P_{\max}^A=20$ dB.}
 	\vspace{-13pt}
 	\label{fig:alpha_R}
 \end{figure}
 
 Figure ~\ref{fig:alpha_R} illustrates the effect of the maximum amplification factor $\Gamma_{\max}$ on the achievable rate $R$. As $\Gamma_{\max}$ increases, the $R$ obtained from the active RIS scheme first grows and then gradually flattens out. This is due to the growing $\Gamma_{\max}$, which gradually diminishes the constraints on the amplification function of the active RIS while the influence of other factors prevails. In addition, the performance differences between the various schemes consistent with those presented in Fig.~\ref{fig:P_BS_EE}, reinforcing the conclusion that MC effect is an indispensable factor in the optimization process of MIMO systems.

%
%
%

\section{Conclusions}\label{sec:Conclusion}

 In this paper, we developed a novel EM-compliant model with MC for active RIS-assisted MIMO wireless system by combining our derived $S$-parameter based active RIS model and multiport network theory, which facilitates system-level analysis and optimization. In order to evaluate the performance of the EM-compliant active RIS model with MC, we formulated joint the transmit beamforming and the reflection coefficient optimization problems to maximize the achievable rate subject to power budgets at the transmitter and active RIS. The Sherman-Morrison inversion and Neumann series (SMaN) based alternating optimization (AO) algorithm was efficiently developed to solve the nonconvex problem, aiming to maximize the achievable rate of the system, specifically the MC effect, are crucial in the optimization process of MIMO systems. Neglecting this effect can result in a substantial performance deficit.

\begin{appendices}
\section{Proof of Theorem 1}
 
 To begin with, based on Eqs.~(\ref{S_parameter})-(\ref{S_parameter3}), we can get
 \begin{align}\label{b_T}
 \boldsymbol{\rm b}_T=\boldsymbol{\rm S}_{TT}\boldsymbol{\rm a}_T+\boldsymbol{\rm S}_{TA}\boldsymbol{\rm a}_A+\boldsymbol{\rm S}_{TR}\boldsymbol{\rm a}_R,\\
 \boldsymbol{\rm b}_A=\boldsymbol{\rm S}_{AT}\boldsymbol{\rm a}_T+\boldsymbol{\rm S}_{AA}\boldsymbol{\rm a}_A+\boldsymbol{\rm S}_{AR}\boldsymbol{\rm a}_R,\label{b_A}\\
 \boldsymbol{\rm b}_R=\boldsymbol{\rm S}_{RT}\boldsymbol{\rm a}_T+\boldsymbol{\rm S}_{RA}\boldsymbol{\rm a}_A+\boldsymbol{\rm S}_{RR}\boldsymbol{\rm a}_R\label{b_R}.
 \end{align}
 Then, substituting Eq.~(\ref{a_A}) into (\ref{b_A}) yields
 \begin{align}\label{a_A_dairu}
 \begin{split}
 \boldsymbol{\rm a}_A=&\boldsymbol{\rm \Gamma}_{A}\left(\boldsymbol{\rm I}_{M}-\boldsymbol{\rm \Gamma}_{A}\boldsymbol{\rm  S}_{AA}\right)^{-1}\left(\boldsymbol{\rm S}_{AT}\boldsymbol{\rm a}_T+\boldsymbol{\rm S}_{AR}\boldsymbol{\rm a}_R\right)\\
 &+\boldsymbol{\rm \Gamma}_{A}\left(\boldsymbol{\rm I}_{M}-\boldsymbol{\rm \Gamma}_{A}\boldsymbol{\rm  S}_{AA}\right)^{-1}\boldsymbol{\rm a}_N.
 \end{split}
 \end{align}
 Afterwards, by inserting Eq.~(\ref{a_A_dairu}) into (\ref{b_T}) and (\ref{b_R}), we can get
 \begin{align}\label{b_T_dairu}
 \boldsymbol{\rm b}_T=\boldsymbol{\rm \widetilde S}_{TT}\boldsymbol{\rm a}_T+\boldsymbol{\rm \widetilde S}_{TR}\boldsymbol{\rm a}_R+\boldsymbol{\rm S}_{TA}\boldsymbol{\rm \Gamma}_{A}\left(\boldsymbol{\rm I}_{M}-\boldsymbol{\rm \Gamma}_{A}\boldsymbol{\rm  S}_{AA}\right)^{-1}\boldsymbol{\rm a}_N,\\ \boldsymbol{\rm b}_R=\boldsymbol{\rm \widetilde S}_{RT}\boldsymbol{\rm a}_T+\boldsymbol{\rm \widetilde S}_{RR}\boldsymbol{\rm a}_R+\boldsymbol{\rm S}_{RA}\boldsymbol{\rm \Gamma}_{A}\left(\boldsymbol{\rm I}_{M}-\boldsymbol{\rm \Gamma}_{A}\boldsymbol{\rm  S}_{AA}\right)^{-1}\boldsymbol{\rm a}_N.\label{b_R_dairu}
 \end{align}
 
 Substituting Eq.~(\ref{a_R}) into (\ref{b_R_dairu}) yields
 \begin{align}\label{b_R_dairu1}
 \boldsymbol{\rm b}_R=\left(\boldsymbol{\rm I}_{N_R}-\boldsymbol{\rm \widetilde S}_{RR}\boldsymbol{\rm \Gamma}_{R}\right)^{-1}\boldsymbol{\rm \widetilde S}_{RT}\boldsymbol{\rm a}_T+\boldsymbol{\rm S}_{N}\boldsymbol{\rm a}_N.
 \end{align}
 Next, associating Eq.~(\ref{b_R_dairu1}) and (\ref{b_R_dairu}), we obtain
 \begin{align}\label{a_R_dairu}
 \boldsymbol{\rm a}_R=&\left(\boldsymbol{\rm \widetilde S}_{RR}^{-1}\left(\boldsymbol{\rm I}_{N_R}-\boldsymbol{\rm \widetilde S}_{RR}\boldsymbol{\rm \Gamma}_{R}\right)^{-1}\boldsymbol{\rm \widetilde S}_{RT}-\boldsymbol{\rm \widetilde S}_{RR}^{-1}\boldsymbol{\rm \widetilde S}_{RT}\right)\boldsymbol{\rm a}_T\notag\\
 &+\boldsymbol{\rm \widetilde S}_{RR}^{-1}\boldsymbol{\rm \widetilde S}_{N}\boldsymbol{\rm a}_N.
 \end{align}
 By inserting Eq.~(\ref{a_R_dairu}) into (\ref{b_T_dairu}), we have
 \begin{align}\label{b_T_dairu1}
 \boldsymbol{\rm b}_T=\boldsymbol{\rm \widehat S}_{TT}\boldsymbol{\rm a}_T+\boldsymbol{\rm \widehat S}_{N}\boldsymbol{\rm a}_N.
 \end{align}
 Then, by substituting Eq.~(\ref{a_T}) into (\ref{b_T_dairu1}), we can get the relationship among $\boldsymbol{\rm a}_T$, $\boldsymbol{\rm a}_S$, and $\boldsymbol{\rm a}_N$ as follows:
 \begin{align}\label{a_T_dairu}
 \boldsymbol{\rm a}_T=\left(\boldsymbol{\rm I}_{N_T}-\boldsymbol{\rm \Gamma}_{T}\boldsymbol{\rm \widehat S}_{TT}\right)^{-1}\hspace{-3mm}\boldsymbol{\rm a}_S+\left(\boldsymbol{\rm I}_{N_T}-\boldsymbol{\rm \Gamma}_{T}\boldsymbol{\rm \widehat S}_{TT}\right)^{-1}\hspace{-2mm}\boldsymbol{\rm \Gamma}_{T}\boldsymbol{\rm \widehat S}_{N}\boldsymbol{\rm a}_N.
 \end{align}
 Finally, inserting Eq.~(\ref{a_T_dairu}) into (\ref{b_R_dairu1}), we can obtain the EM-compliant model for active RIS-assisted MIMO communication, ie., Eq.~(\ref{channel_model1}). Proof is complete.

 \section{Proof of Theorem 2}
 
 To prove Theorem~\ref{theorem_2}, we resort to the process of proving Theorem~\ref{theorem_channnel}. Substituting Eqs.~(\ref{a_T_dairu}) into (\ref{a_A_dairu}) yields
 \begin{align}\label{a_A_dairu1}
 \boldsymbol{\rm a}_A\hspace{-1mm}=&\boldsymbol{\rm \Gamma}_{A}\hspace{-1mm}\left(\boldsymbol{\rm I}_{M}\hspace{-1mm}-\hspace{-1mm}\boldsymbol{\rm \Gamma}_{A}\boldsymbol{\rm  S}_{AA}\hspace{-1mm}\right)^{-1}\hspace{-1mm}\boldsymbol{\rm S}_{AT}\hspace{-1mm}\left(\hspace{-1mm}\boldsymbol{\rm I}_{N_T}\hspace{-1mm}-\hspace{-1mm}\boldsymbol{\rm \Gamma}_{T}\boldsymbol{\rm \widehat S}_{TT}\hspace{-1mm}\right)^{-1}\hspace{-2mm}\left(\hspace{-1mm}\boldsymbol{\rm a}_S+\boldsymbol{\rm \Gamma}_{T}\boldsymbol{\rm \widehat S}_{N}\boldsymbol{\rm a}_N\hspace{-1mm}\right)\nonumber\\
 &+\boldsymbol{\rm \Gamma}_{A}\hspace{-1mm}\left(\boldsymbol{\rm I}_{M}\hspace{-1mm}-\hspace{-1mm}\boldsymbol{\rm \Gamma}_{A}\boldsymbol{\rm  S}_{AA}\hspace{-1mm}\right)^{-1}\boldsymbol{\rm S}_{AR}\boldsymbol{\rm a}_R+\boldsymbol{\rm \Gamma}_{A}\hspace{-1mm}\left(\boldsymbol{\rm I}_{M}\hspace{-1mm}-\hspace{-1mm}\boldsymbol{\rm \Gamma}_{A}\boldsymbol{\rm  S}_{AA}\hspace{-1mm}\right)^{-1}\hspace{-1mm}\boldsymbol{\rm a}_N.
 \end{align}
 Then, inserting Eq.~(\ref{channel_model1}) into (\ref{a_A_dairu1}) and applying the unilateral approximation, we can obtain
  \begin{align}\label{a_A_dairu2}
 \boldsymbol{\rm a}_A\hspace{-1mm}=&\boldsymbol{\rm \Gamma}_{A}\hspace{-1mm}\left(\boldsymbol{\rm I}_{M}\hspace{-1mm}-\hspace{-1mm}\boldsymbol{\rm \Gamma}_{A}\boldsymbol{\rm  S}_{AA}\hspace{-1mm}\right)^{-1}\hspace{-1mm}\boldsymbol{\rm S}_{AT}\hspace{-1mm}\left(\hspace{-1mm}\boldsymbol{\rm I}_{N_T}\hspace{-1mm}-\hspace{-1mm}\boldsymbol{\rm \Gamma}_{T}\boldsymbol{\rm \widehat S}_{TT}\hspace{-1mm}\right)^{-1}\hspace{-2mm}\left(\hspace{-1mm}\boldsymbol{\rm a}_S+\boldsymbol{\rm \Gamma}_{T}\boldsymbol{\rm \widehat S}_{N}\boldsymbol{\rm a}_N\hspace{-1mm}\right)\nonumber\\
 &+\boldsymbol{\rm \Gamma}_{A}\hspace{-1mm}\left(\boldsymbol{\rm I}_{M}\hspace{-1mm}-\hspace{-1mm}\boldsymbol{\rm \Gamma}_{A}\boldsymbol{\rm  S}_{AA}\hspace{-1mm}\right)^{-1}\hspace{-1mm}\boldsymbol{\rm a}_N.
 \end{align}
 
 Finally, by substituting $\boldsymbol{\rm \Gamma}_{T} = 0$ and $\boldsymbol{\rm \Gamma}_{R} = 0$ into (\ref{a_A_dairu2}), we can derive the relationship among $\boldsymbol{\rm a}_T$, $\boldsymbol{\rm a}_S$, and $\boldsymbol{\rm a}_N$ as presented in Eq.~(\ref{theorem_21}). Proof is complete.

\end{appendices}
 
\bibliographystyle{IEEEtran}
\bibliography{References}

\end{document}